\documentclass[prb,twocolumn,epsfig,epsf]{revtex4}
\usepackage{amsmath,amsthm,amsfonts,amscd,mathrsfs,pifont,bm}
\usepackage{eucal,graphicx,color}
\usepackage{esint}

\makeatletter
\@for\next:={int,iint,iiint,iiiint,dotsint,oint,oiint,sqint,sqiint,
  ointctrclockwise,ointclockwise,varointclockwise,varointctrclockwise,
  fint,varoiint,landupint,landdownint}\do{%
    \expandafter\edef\csname\next\endcsname{%
      \noexpand\DOTSI
      \expandafter\noexpand\csname\next op\endcsname
      \noexpand\ilimits@
    }%
  }
\makeatother

\newcommand{\beq}{\begin{equation}}
\newcommand{\eeq}{\end{equation}}

\renewcommand{\Re}{\mathrm{Re}}
\renewcommand{\Im}{\mathrm{Im}}

\newcommand{\M}[1]{\pmb{#1}}

\begin{document}

\title  {Dynamics of Disordered Mechanical Systems with Large Connectivity, Free Probability Theory, and Quasi-Hermitian Random Matrices}

%\author {Joshua Feinberg\orcidJosh\footnote{{\includegraphics[scale=0.7]{ORCID.png}} https://orcid.org/0000-0002-2869-0010} and Roman Riser}
\author {Joshua Feinberg\footnote{ https://orcid.org/0000-0002-2869-0010} and Roman Riser}

\affiliation
       {Department of Mathematics\\ and\\
Haifa Research Center for Theoretical Physics and Astrophysics University of Haifa, Haifa 31905, Israel}

\begin  {abstract}
Disordered mechanical systems with high connectivity represent a limit opposite to the more familiar case of disordered crystals. Individual ions in a crystal are subjected essentially to nearest-neighbor interactions. In contrast, the systems studied in this paper have all their degrees of freedom coupled to each other. Thus, the problem of linearized small oscillations of such systems involves two full positive-definite and non-commuting matrices, as opposed to the sparse matrices associated with disordered crystals. Consequently, the familiar methods for determining the averaged vibrational spectra of disordered crystals, introduced many years ago by Dyson and Schmidt, are inapplicable for highly connected disordered systems. In this paper we apply random matrix theory (RMT) to calculate the averaged vibrational spectra of such systems, in the limit of infinitely large system size. At the heart of our analysis lies a calculation of the average spectrum of the product of two positive definite random matrices by means of free probability theory techniques.  We also show that this problem is intimately related with {\em quasi-hermitian random matrix theory} (QHRMT), which means that the `hamiltonian' matrix is hermitian with respect to a non-trivial metric. This extends ordinary hermitian matrices, for which the metric is simply the unit matrix. The analytical results we obtain for the spectrum agree well with our numerical results. The latter also exhibit oscillations at the high-frequency band edge, which fit well the Airy kernel pattern. We also compute inverse participation ratios of the corresponding amplitude eigenvectors and demonstrate that they are all extended, in contrast with conventional disordered crystals. Finally, we compute the thermodynamic properties of the system from its spectrum of vibrations.  In addition to matrix model analysis, we also study the vibrational spectra of various multi-segmented disordered pendula, as concrete realizations of highly connected mechanical systems.  A universal feature of the density of vibration modes, common to both pendula and the matrix model, is that it tends to a non-zero constant at vanishing frequency.
\end{abstract}

%\pacs   {73.23.--b, 05.45.Mt, 02.30.Ik}
\maketitle
\newpage
\section{Introduction}\label{introduction}
\subsection{Small Oscillations}\label{small oscillations}
The problem of determining the small oscillations of a mechanical system about a stable equilibrium state is ubiquitous in physics. Thus, given a system with $N$ degrees of freedom and corresponding  generalized coordinates ${\bf q} = (q_1,\ldots q_N)$, its small oscillations about a stable equilibrium point ${\bf q}_0$ are solutions of the linearized equations of motion \cite{LL, Arnold, GK}
\begin{equation}\label{small-oscillations}
\M{M} \ddot {\bf x} +  \M{K} {\bf x}  = 0\,, 
\end{equation}
where the $N\times N$ {\em strictly positive-definite} matrix $\M{M} = \M{a}({\bf q}_0)$ is the value of the metric appearing in the kinetic part the lagrangian $L = \frac{1}{2}\dot{\bf q}^T \M{a}({\bf q}) \dot{\bf q} - U({\bf q})$ evaluated at ${\bf q}_0$, the {\em positive} matrix $K_{ij} = {\partial^2 U\over \partial q_i\partial q_j}({\bf q}_0),(i,j=1,2\ldots N)$ is the Hessian of the potential at the equilibrium point, and ${\bf x} = {\bf q}-{\bf q}_0$ is a small deviation from equilibrium. Harmonic eigenmodes of the system are solutions of the form ${\bf x}(t) = {\bf A} e^{i\omega t}$, for which \eqref{small-oscillations} implies the characteristic equation 
\begin{equation}\label{eigen}
\left(-\omega^2 \M{M} + \M{K}\right){\bf A} = {\bf 0}
\end{equation}
for the eigenvector amplitude and frequency eigenvalue. 
Eigenfrequencies are roots of the characteristic polynomial $P_N(\omega^2)= \det \left(-\omega^2 \M{M} + \M{K}\right)$. These roots are all positive, since according to \eqref{eigen}, $\omega^2 ={{\bf A}^\dagger \M{K}\bf{A}\over {\bf A}^\dagger \M{M}{\bf A}}$ is the ratio of two positive quantities. This should be expected on physical grounds, since in the absence of dissipation, the eigenfrequencies $\omega$ of a stable system are all real.

\subsection{Quasi-Hermitian Matrices}\label{quasi-hermitian matrices.}
We can rewrite the eigenmode equation \eqref{eigen} as $\M{H}{\bf A} = \omega^2 {\bf A}$, where the ``hamiltonian" matrix\footnote{$\M{H}$ is the matrix to be diagonalized in order to obtain the eigenfrequencies. Thus, in the parlance of RMT we refer to it as the ``hamiltonian". It is not to be confused, of course, with the hamiltonian \eqref{hamiltonian} of the mechanical system.} is
\begin{equation}\label{H}
\M{H} = \M{M}^{-1}\M{K}\,\quad{\rm and}\quad \M{H}^\dagger = \M{K}\M{M}^{-1}\,.
\end{equation}
In general, $[\M{M},\M{K}]\neq 0$, and consequently $\M{H}^\dagger\neq \M{H}$. However, $\M{H}$ and its adjoint fulfil the {\em intertwining relation}
\begin{equation}\label{intertwining}
\M{H}^\dagger \M{M} = \M{M}\M{H} \quad (=\M{K})\,. 
\end{equation} 
This relation implies that $\M{H}$ is hermitian in a vector space endowed with a non-trivial metric $\M{M}$, namely, 
\begin{equation}\label{metric}
\langle {\bf A}_1 | \M{H}{\bf A}_2\rangle_M  =  \langle \M{H}{\bf A}_1 | {\bf A}_2\rangle_M\,, 
\end{equation}
with inner product $\langle {\bf A}_1 | {\bf A}_2\rangle_M = \langle {\bf A}_1 | \M{M}{\bf A}_2\rangle = {\bf A}^\dagger_1\M{M}{\bf A}_2$ (and where, of course, $\langle {\bf A}_1 | {\bf A}_2\rangle =  {\bf A}^\dagger_1{\bf A}_2$ is the standard inner product, corresponding to $\M{M}=\M{1}$, with respect to which the adjoint $\M{H}^\dagger$ in (\ref{intertwining}) is defined). The intertwining relation \eqref{intertwining}
is equivalent to the similarity transformation
\begin{equation}\label{similarity}
\M{H}^\dagger  = \M{M}\M{H}\M{M}^{-1}
\end{equation}
between $\M{H}$ and its adjoint.  A corollary of this is that the characteristic polynomial of $\M{H}$ has real coefficients: $\left(\det (z-\M{H})\right)^* = \det (z^*-\M{H})\,,$ consistent with the fact that all eigenvalues of $\M{H}$ are real.   
Another way to establish positivity of the eigenvalues of $\M{H}$ is to observe from \eqref{intertwining} (or  \eqref{similarity})  that both $\M{H}$ and $\M{H}^\dagger$ are similar to a positive hermitian matrix, namely, 
\begin{equation}\label{similarity1}
\M{H} = \frac{1}{\sqrt{\M{M}}} \M{h}\sqrt{\M{M}}\,,\quad {\rm and}\quad  \M{H}^\dagger = \sqrt{\M{M}} \M{h}\frac{1}{\sqrt{\M{M}}}\,,
\end{equation}
where we chose $\sqrt{\M{M}}$ as the positive definite square root of $\M{M}$, and 
\begin{equation}\label{h}
\M{h} = \frac{1}{\sqrt{\M{M}}} \M{K} \frac{1}{\sqrt{\M{M}}} 
\end{equation}
is manifestly hermitian and positive. The matrix $\M{H}$ is said to be\cite{talks} a {\em strictly quasi-hermitian} matrix, due to its similarity \eqref{similarity1} to the hermitian matrix $\M{h}$. 
The similarity matrix $\M{S} = \sqrt{\M{M}}$ in (\ref{similarity1}) is hermitian as well. This need not be the case in general: A matrix $\M{H}$ is said to be strictly quasi-hermitian (sQH), if it is similar to a hermitian matrix $\M{h}$ 
\begin{equation}\label{similarity2}
\M{H} = \M{S}^{-1} \M{h}\M{S}\,,\quad {\rm and}\quad  \M{H}^\dagger = \M{S}^\dagger \M{h}\M{S}^{\dagger^{-1}}\,,
\end{equation}
with $\M{S}$ a complex invertible matrix. Thus, an sQH matrix $\M{H}$ is diagonalizable, and all its eigenvalues are real. Moreover, it fulfils the intertwining relation 
\begin{equation}\label{intertwining1}
\M{H}^\dagger \left(\M{S}^\dagger \M{S}\right) = \left(\M{S}^\dagger \M{S}\right) \M{H}\,,
\end{equation} 
which means that $\M{H}$ is hermitian with respect to the metric $\M{M} = \M{S}^\dagger \M{S}$. 

If invertibility of the metric $\M{M}$ is relaxed, then $\M{H}$ is merely a {\em quasi-hermitian} (QH) matrix. (For a useful clarification of terminology see \cite{Fring-Assis}. In this paper we introduce sQH matrices and make explicit distinction between them and QH matrices.)

One can also consider {\it sQH random matrix models} (see also Section \ref{QHRMT}). An interesting sQH random matrix model was introduced in \cite{Joglekar}. These authors used the fact that {\em given} a metric $\M{M}=\M{S}^\dagger \M{S}$, then considering the intertwining relation (\ref{intertwining1}) as an equation for $\M{H}$, its solution is  
\begin{equation}\label{intertwining-solution}
\M{H} = \M{A}\M{M}\,, \quad{\rm for~any}\quad \M{A} = \M{A}^\dagger\,.
\end{equation}
Thus, given $\M{M}$, the linear homogeneous equation \eqref{intertwining1} has $N^2$ independent solutions for $\M{H}$. The authors of \cite{Joglekar} then fixed a metric, and took the hamiltonian $\M{H}$ as random, with the aim of studying numerically the dependence of the average density of eigenvalues and level spacing statistics on the metric. Yet another interesting example of a sQH random matrix model, akin to the Dicke model of superradiance, was provided by \cite{Deguchi}, in which a numerical study of the level spacing distribution was carried out. 

It follows from \eqref{intertwining-solution} that the eigenvalue problem $\M{H}\M{u} = \lambda\M{u}$ for the sQH matrix $\M{H}$ is equivalent to\begin{equation}\label{pencil}
(\M{\tilde A}-\lambda\M{M})\M{u} = 0\,,
\end{equation}
where we have introduced the hermitian matrix $\M{\tilde A} = \M{M}\M{A}\M{M}$. (In the particular case corresponding to \eqref{eigen} we have, of course, $\M{\tilde A} = \M{K}$ and $\lambda=\omega^2$.) The combination $\M{\tilde A}-\lambda\M{M}$ in \eqref{pencil} constitutes what is known in the mathematical literature as a regular pencil of matrices (or a regular pencil of quadratic forms)\cite{GK, Gantmacher}. The qualifier {\em regular} means here that  $\M{M}$ and  $\M{\tilde A}$ are square matrices, and that $\det (\M{\tilde A}-\lambda\M{M})$ does not vanish identically.  Thus, eigenvalue problems for sQH matrices can always be associated with regular matrix pencils.

Quasi-hermitian matrices can be thought of as truncated quasi-hermitian linear operators. For an early consideration of quasi-hermitian operators (in which this term was coined) see \cite{Dieudonne}. For general considerations on the construction of consistent quantum mechanical systems based on a QH hamiltonian and observables see \cite{stellenbosch}. Such QH quantum mechanical systems are closely related to ${\cal PT}$-symmetric quantum mechanical systems \cite{BB,CMB} with unbroken ${\cal PT}$ symmetry. The latter possess real energy spectra due to the existence of a positive-definite metric known as the ${\cal CPT}$ inner product. When the metric operator ceases to be positive-definite, pairs of complex-conjugate eigenvalues appear in the spectrum of the hamiltonian, a situation referred to as broken ${\cal PT}$-symmetry. 
Operators satisfying the intertwining relation $\M{H}^\dagger \M{M} = \M{M}\M{H}$ with indefinite metric $\M{M}$ are sometimes referred to as {\em pseudo-hermitian} (PH) operators \cite{Froissart}. The modern evocation of PH operators was made by\cite{Mostafazadeh}.  For a very brief but useful summary of the history of quasi- and pseudo-hermiticity see \cite{Fring-Assis, Fring}.  

Upon truncation to finite vector spaces, pseudo-hermitian operators turn into pseudo-hermitian matrices. See \cite{Kumar} for a recent discussion of (real asymmetric) pseudo-hermitian random matrices.

\subsection{Plan of the rest of this paper}\label{plan}
In Section \ref{QHRMT} we first put the problem of vibration eigenmodes of highly connected systems in historic context, motivate application of QHRMT to study such systems, and contrast them with analysis of phonons in crystals. Next, as further motivation, we introduce a family of clean (uniform) and disordered multi-segmented pendula as concrete examples of highly connected mechanical systems. We study numerically the spectral statistics and localization properties of these systems in some detail. Only then do we turn to defining our random matrix model, and study it in detail both analytically and numerically. In particular, we obtain an explicit large-$N$ expression for the density of eigenvalues - that is, vibrational eigenfrequencies, and study its universal edge-behavior numerically. We also study its statistics of eigenvectors. An interesting observation is that the density of eigenfrequencies of both our matrix model and the pendula tend to a nonvanishing constant in the limit of small frequencies, which seems to be a common universal feature of highly connected systems.  We have recently reported this universal behavior in \cite{FRDecember}.

In Section \ref{Liuovillian} we explain how the diagrammtic approach of  \cite{BJN} to $S$-transforms in free probability theory can be interpreted in terms of the Liouvillian of our mechanical system. 

Finally, in Section \ref{phonons}  we analyze the thermodynamic properties of phonons in our random matrix model at equilibrium. 

\section{Strictly Quasi-Hermitian Random Matrix Theory for Small Oscillations}\label{QHRMT}
Under certain conditions, the problem of small oscillations \eqref{small-oscillations}-\eqref{eigen} naturally lends itself to analysis by means of random matrices. One very important example is the determination of the average phonon spectrum of disordered crystals. This problem was solved long ago in one-dimension by Dyson \cite{Dyson} and Schmidt \cite{Schmidt}. Individual ions in a disordered chain are subjected to nearest-neighbor interactions. Consequently, the mass matrix $\M{M}$ is diagonal, and the spring-constant matrix $\M{K}$ is tri-diagonal (a Jacobi matrix). Randomness arises either due to having random ion masses on the diagonal of $\M{M}$, or random spring constants in $\M{K}$ (or due to both). 

Disordered mechanical systems with high connectivity represent a limit opposite to the more familiar case of disordered crystals. Such systems have all their degrees of freedom coupled to each other. Thus, the problem of small oscillations in such systems involves two full positive-definite and non-commuting matrices $\M{M}$ and $\M{K}$. Such systems may be inherently random (due to a random metric $\M{a}({\bf q})$ or potential $U({\bf q})$ in the lagrangian), or following Wigner's original introduction of random matrix theory into nuclear physics, may be just approximated by random matrices due to high structural complexity of the system under study. 

\subsection{Multi-Segmented Pendulum}\label{pendulum}
As a concrete physical realization of the latter possibility, consider a multi-segmented pendulum made of $N$ rigid (massless) segments of lengths $l_1, \ldots l_N$ and point masses $m_1, \ldots m_N$.  The mass $m_k$ is attached to the frictionless hinge connecting segments $l_k$ and $l_{k+1}$, and carries electric charge $Q_k$. The charges $Q_1,\ldots Q_N$ are all like-sign, say positive, rendering all Coulomb interactions repulsive. The pendulum is suspended by the other end of the  first segment $l_1$ from a frictionless hinge, which is fixed to an infinite mass $m_0$ (a wall), carrying positive charge $Q_0$. The whole system is suspended in Earth's gravity $g$, and is free to execute planar oscillations. (This system can be thought of as a model for a charged (unscreened) polymer chain in a uniform external field.)

Let $ \theta_k\in[-\pi,\pi]$ be the angle between the segment $l_k$ and the downward vertical. Clearly, the stable equilibrium state of the system occurs when all segments align vertically, i.e.~when all $\theta_k=0$. We have studied the small oscillations of this pendulum about its equilibrium state, i.e.~motions for which all $|\theta_k|\ll 1$. The lagrangian governing these small oscillations is $L=\frac{1}{2} \dot{\theta}^T \M{M} \dot{\theta}-\frac{1}{2}\theta^T \M{K} \theta$, with $\theta=(\theta_1,\ldots \theta_N)^T$, and
where the entries of the symmetric matrices $\M{M}$ and $\M{K}$ are given by
\begin{align*}
M_{ij}&=l_i l_j \!\!\!\!\!\! \sum_{~~~k=\max(i,j)}^{N}\!\!\!\!\!\!m_k,  &K_{ij}&=U_{ij}+\delta_{ij} l_i \,g \sum_{k=i}^{N} m_k,\nonumber\\
U_{ij}&=\left\{ \begin{array}{ll}
-\tilde{U}_{ij}, & \text{if $i\neq j$,}\\ \sum\limits_{\substack{k=1 \\ k\neq i}}^N\tilde{U}_{ik}, & \text{if $i=j$,}
\end{array} \right.\!\!\!\!\!\! \nonumber\\
\tilde{U}_{ij}&=l_i l_j \!\!\!\sum_{k=1}^{~\min(i,j)}  \!\!\!\sum_{~~l=\max(i,j)}^N  \!\!\! Q_{kl}, &Q_{ij}&=Q_{i-1} Q_{j}\left(\sum_{k=i}^{j}l_k \right)^{\!\!\!-3}.
\end{align*}
The matrix $\M{U}$ originates from the Coulomb potential energy. Note that the sum of entries in each row of this matrix vanishes. This is simply a manifestation of translational invariance of the Coulomb interaction: It depends only on squares of differences of angles $(\theta_i-\theta_j)^2$ (which arise from expanding $\cos(\theta_i-\theta_j)$ to the first nontrivial order). Therefore, shifting {\em all} $\theta_i$ by the same amount $\theta_0$ cannot change the Coulomb energy of the system. In particular, a configuration in which all masses align along a straight line making some fixed angle $\theta_0$ with the vertical has the same Coulomb energy as the equilibrium configuration. In other words, 
$\theta=(\theta_0,\ldots \theta_0)^T$ must be a null eigenvector of the matrix $\M{U}$, which is indeed the case. (The gravitational potential energy breaks this translational symmetry, of course.) Moreover, all other eigenvalues of $\M{U}$ should be positive, since any distortion of the pendulum from a straight line configuration will cost energy due to Coulomb repulsion. Indeed, $\M{U}$ is a diagonally dominant matrix, and is therefore positive semi-definite by virtue of Gershgorin's theorem\cite{Gershgorin}. 

If all charges and masses are non-zero, $\M{M}$ and $\M{K}$ are full matrices (the latter is due to the long-range Coulomb interaction), while in the pure gravitational model, for which $Q_0=Q_1=\ldots=Q_N=0$, the matrix $\M{K}$ is diagonal.

\begin{figure}[b]
\includegraphics[width=\columnwidth]{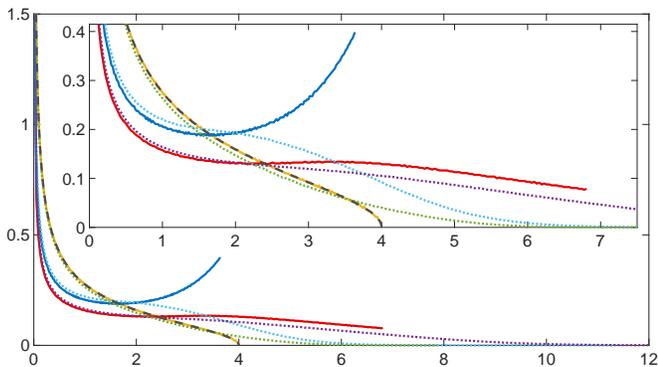}
\caption{Histogram envelope curve showing the density of states (times $N^2$) for the multi-segmented pendulum as a function of $\omega^2/N^2$. Solid lines correspond to uniform systems with all lengths, masses and charges being equal and set to $l_k=m_k=1/N$ and $Q_k={\cal Q}/(N \sqrt{\log N})$ with $N=16384$. The three graphs correspond to ${\cal Q}=1, g=1$ (red), ${\cal Q}=0, g=1$ (yellow), and ${\cal Q}=1, g=0$ (blue). The dotted lines in purple, green and cyan show the analogous curves respectively, when lengths, masses and charges are random and drawn from the  probability distributions defined in the text,  with mean values corresponding to the constant cases, and with $N=1024$ averaged over $25000$ samples. The dashed black line overlying the yellow line shows the Marchenko-Pastur distribution. The inset here (and in all figures below) shows a magnification of the edge behavior for the same data.}\label{fig:pendulum}
\end{figure}

In order to get oriented, let us consider first the uniform pendulum, for which all lengths, masses and charges are equal. As we want our system to have finite total length $L$ and mass $M$ when $N\rightarrow \infty$, segment lengths and point masses must scale like $l_k\sim N^{-1}$ and $m_k\sim N^{-1}$. Under such circumstances the diagonal elements $U_{ii}$ (for segments $1 \ll i \ll N $ in the bulk of the pendulum) grow like $\log N =  \log (L/l)$, because it is essentially the electrostatic potential at a point somewhere on a uniformly charged rod  of length $L$\cite{Sommerfeld}. Therefore, we have to scale the charges like $Q_k\sim (N \sqrt{\log N})^{-1}$, in order to balance the Coulomb potential energy against gravitational potential energy. Under this scaling of parameters, careful inspection of the matrices $\M{M}$ and $\M{K}$ shows that $\M{H} = \M{M}^{-1}\M{K}$ scales like $N^2$, rendering its eigenvalue bandwidth scaling likewise. We have investigated such a system numerically. The solid lines in Fig.~\ref{fig:pendulum} represent histogram envelopes for $\varrho_{\M{H}}(\omega^2)$, the (normalized) density of states (eigenvalues of $\M{H}$), defined in Eq.(\ref{Hdensity}), as a function of $\omega^2/N^2$. These lines correspond to the cases of a mixed system with both Coulomb and gravitational interaction, a system with purely Coulomb interaction, and a system with purely gravitational interaction.  
Remarkably, as $\omega\rightarrow 0$, the density $\varrho_{\M{H}}(\omega^2)$ in all three cases diverges universally as 
\begin{equation}\label{universal}
\varrho_{\M{H}}(\omega^2) \sim {c\over\sqrt{\omega^2}} = {c\over\omega}
\end{equation} 
(with coefficients $c\neq 0$ which vary from case to case). This means that the density of frequency eigenmodes $\tilde\rho(\omega)  = 2\omega \varrho_{\M{H}}(\omega^2)$, defined in Eq.\eqref{mode-density}, tends to a constant $\tilde\rho(0) = 2c$ in this limit. 

The spectrum of the mixed system is noticeably broader due to the combined Coulomb and gravitational forces acting on the masses. While the density of states in the purely gravitational model  vanishes at the right edge of the spectrum like a square root, it exhibits a band-end discontinuity when Coulomb interactions are involved. Such discontinuous behavior of the density of states occurs also in other and completely different physical systems, such as Bloch electrons in a perfect crystal. 

Finally, an utterly surprising observation, demonstrated by the coincidence of the yellow and black dashed lines in Fig.~\ref{fig:pendulum}, is that the density of eigenvalues of the purely gravitational and completely deterministic system follows the Marchenko-Pastur distribution \cite{Marchenko-Pastur} given by \eqref{density} below.

We have also studied the disordered pendulum, with lengths $l_k$, masses $m_k$ and charges $Q_k$ all being i.i.d. random variables, drawn from probability distributions chosen such that the mean values of these variables would coincide with the corresponding values of these parameters in the uniform, undisordered systems displayed in Fig.~\ref{fig:pendulum}, and with standard deviations $\sigma$ of the same order of magnitude at large-$N$ as those averages. Specifically, the $l_k$s were taken from the uniform distribution on the interval $(0.8/N,1.2/N)$, the $m_k$s from the uniform distribution on the interval $(0.5/N,1.5/N)$, and the $Q_k$s were taken from a discrete distribution with equiprobable values $0.5/N\sqrt{\log N}$ and $1.5/N\sqrt{\log N}$. The resulting averaged densities of states are displayed by the dotted lines in Fig.~\ref{fig:pendulum}. 
Note that these densities of the disordered systems vanish at the high-frequency edge of the spectrum for all three cases. The band-end spectral discontinuities of some of the ordered uniform systems are smoothed out as a result of averaging over disorder. 

More importantly, note that disorder does not change the universal low-frequency divergence \eqref{universal}. Evidently, low-frequency oscillations correspond to long-wavelength collective motions, which probe the large-scale structure of the pendulum, averaged over many random segments.  Spectra of the disordered systems start to deviate from their uniform system counterparts only as the frequency increases, and oscillations become sensitive to the smaller scale structure of the system.

\begin{figure}[b!]
	\includegraphics[width=\columnwidth]{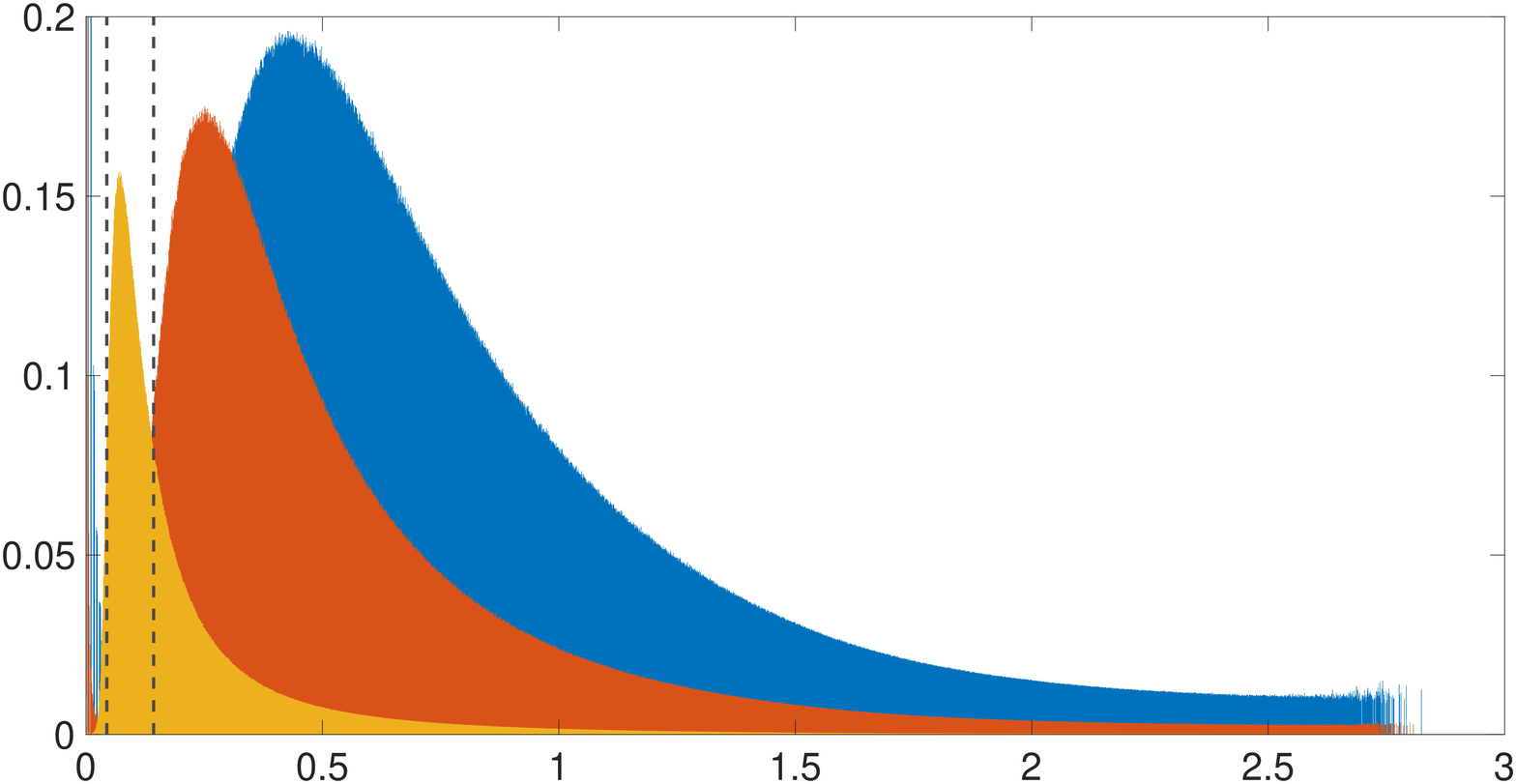}
	\includegraphics[width=\columnwidth]{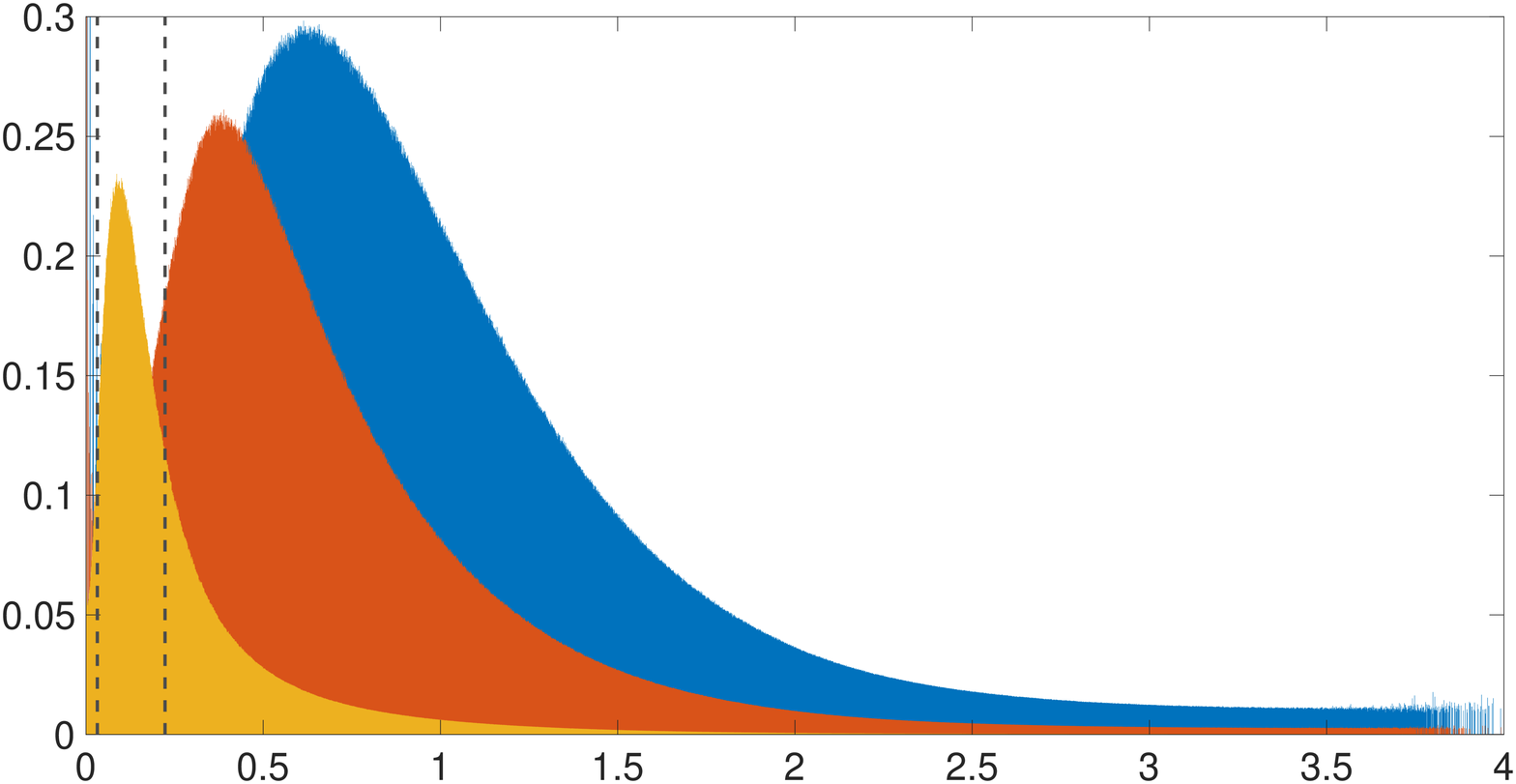}
	\includegraphics[width=\columnwidth]{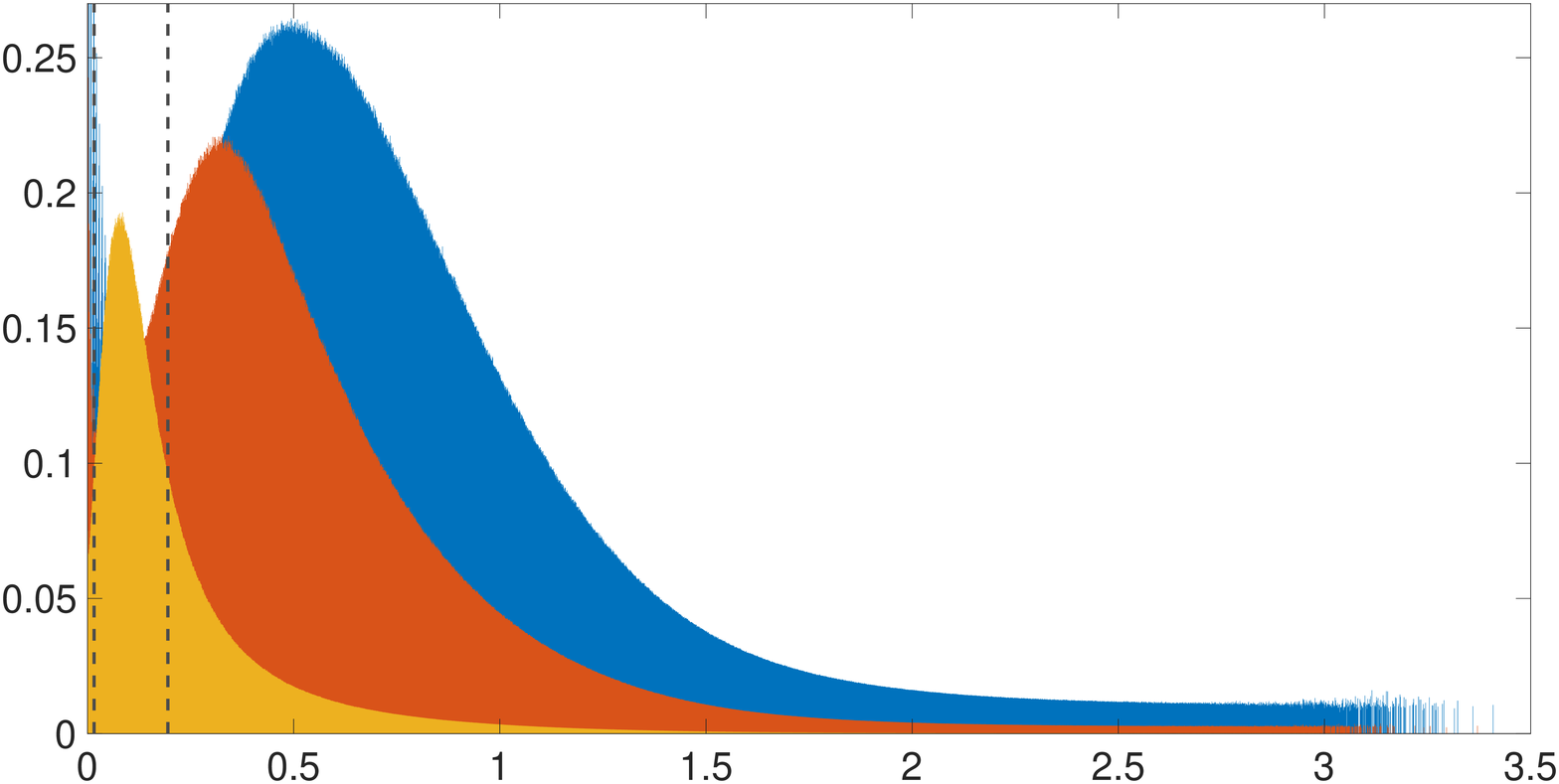}
	\caption{Normalized participation ratios of the three types of disordered pendula: pure gravity (top), pure Coulomb (middle), mixed (bottom). The histograms show the normalized participation ratio for $N=256$ (blue), 1024 (brown), 16384 (yellow), computed from 100000, 25000 and 2000 samples, respectively. The two vertical dashed lines show the full-width-half-maximum of the yellow main peak, which are used in Fig.~\ref{fig:pspend} to separate the bandwidth into different regions.}\label{fig:prpend}
\end{figure}
\begin{figure}[b!]
	\includegraphics[width=\columnwidth]{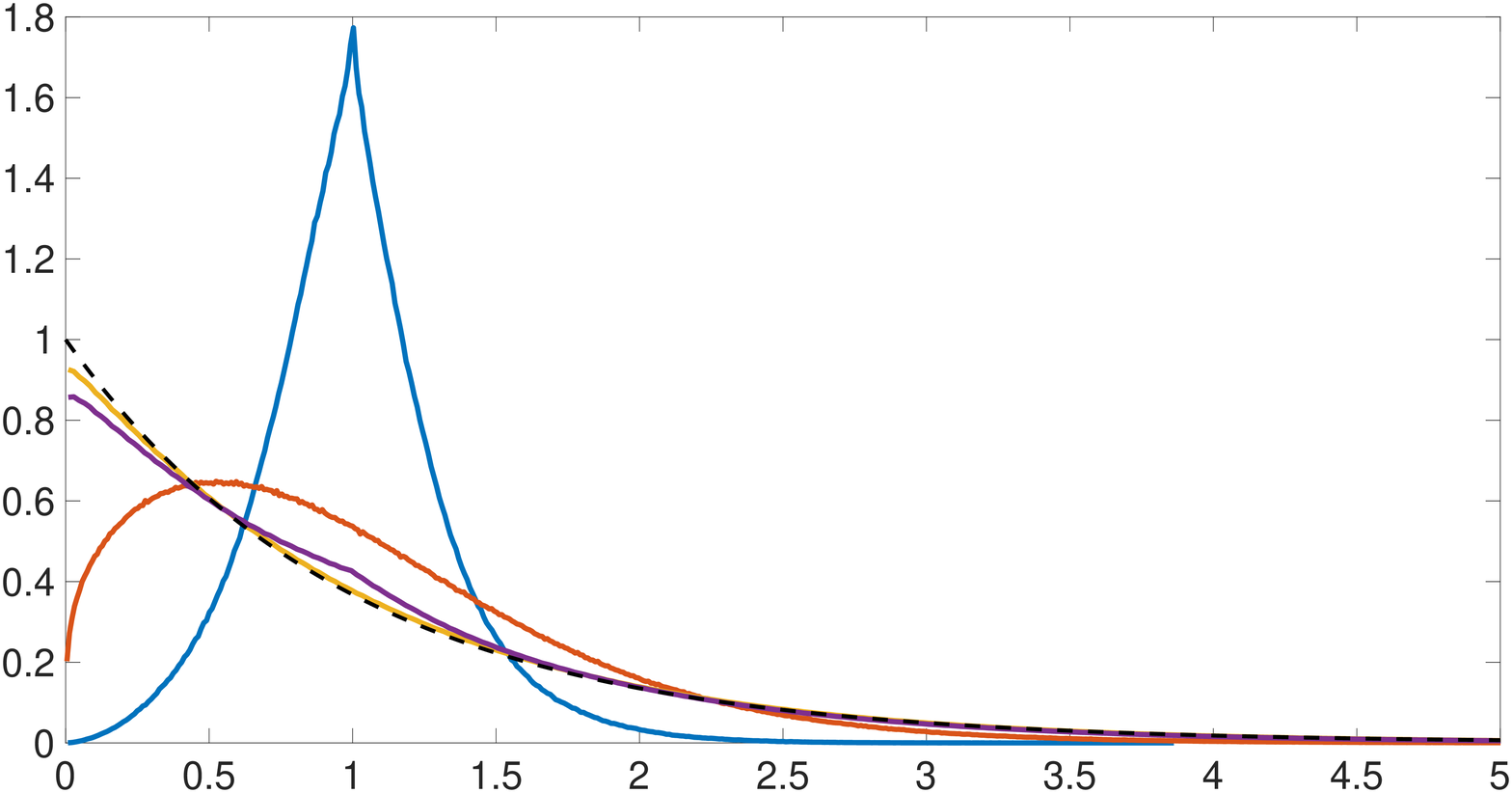}
	\includegraphics[width=\columnwidth]{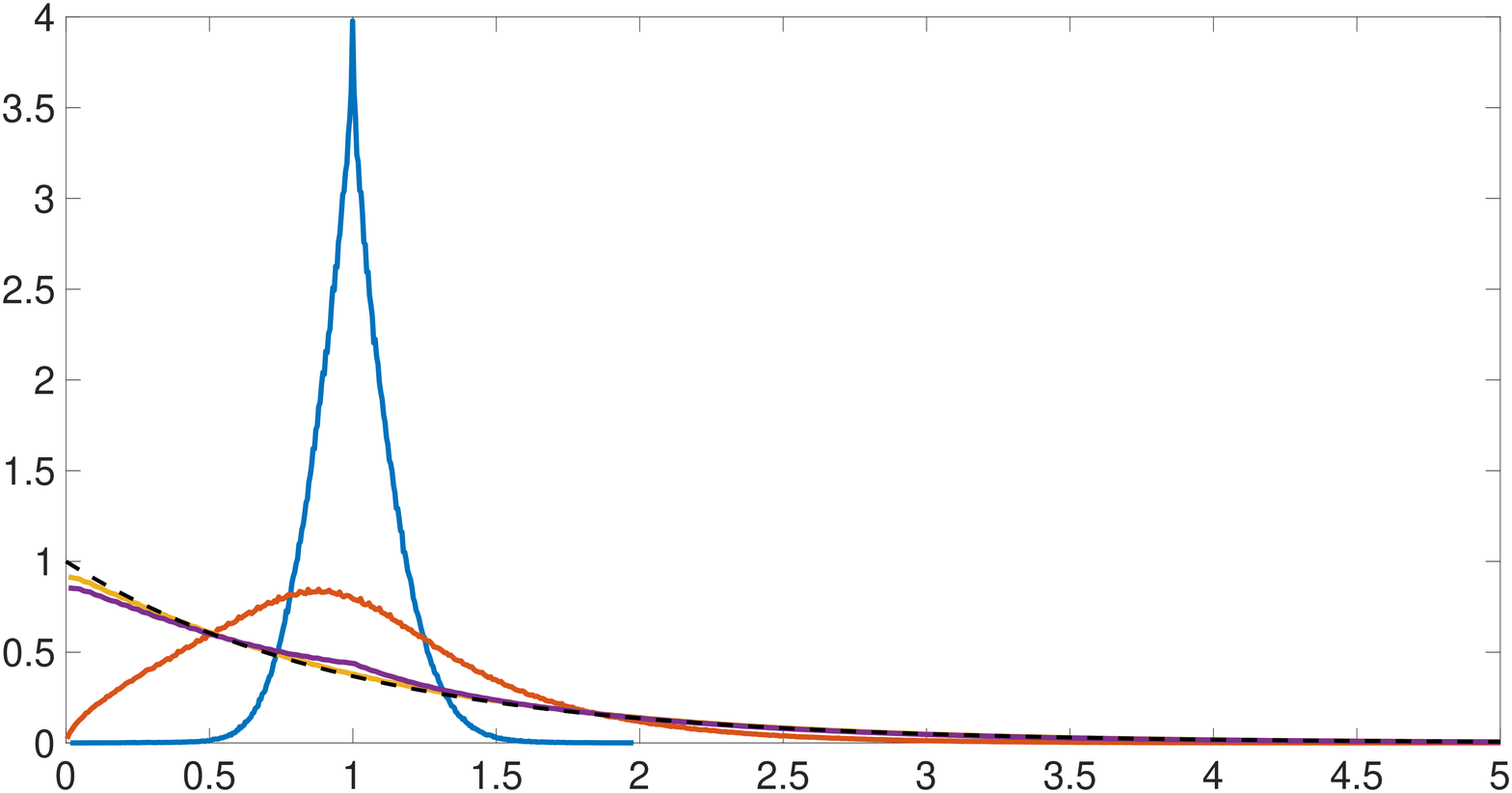}
	\includegraphics[width=\columnwidth]{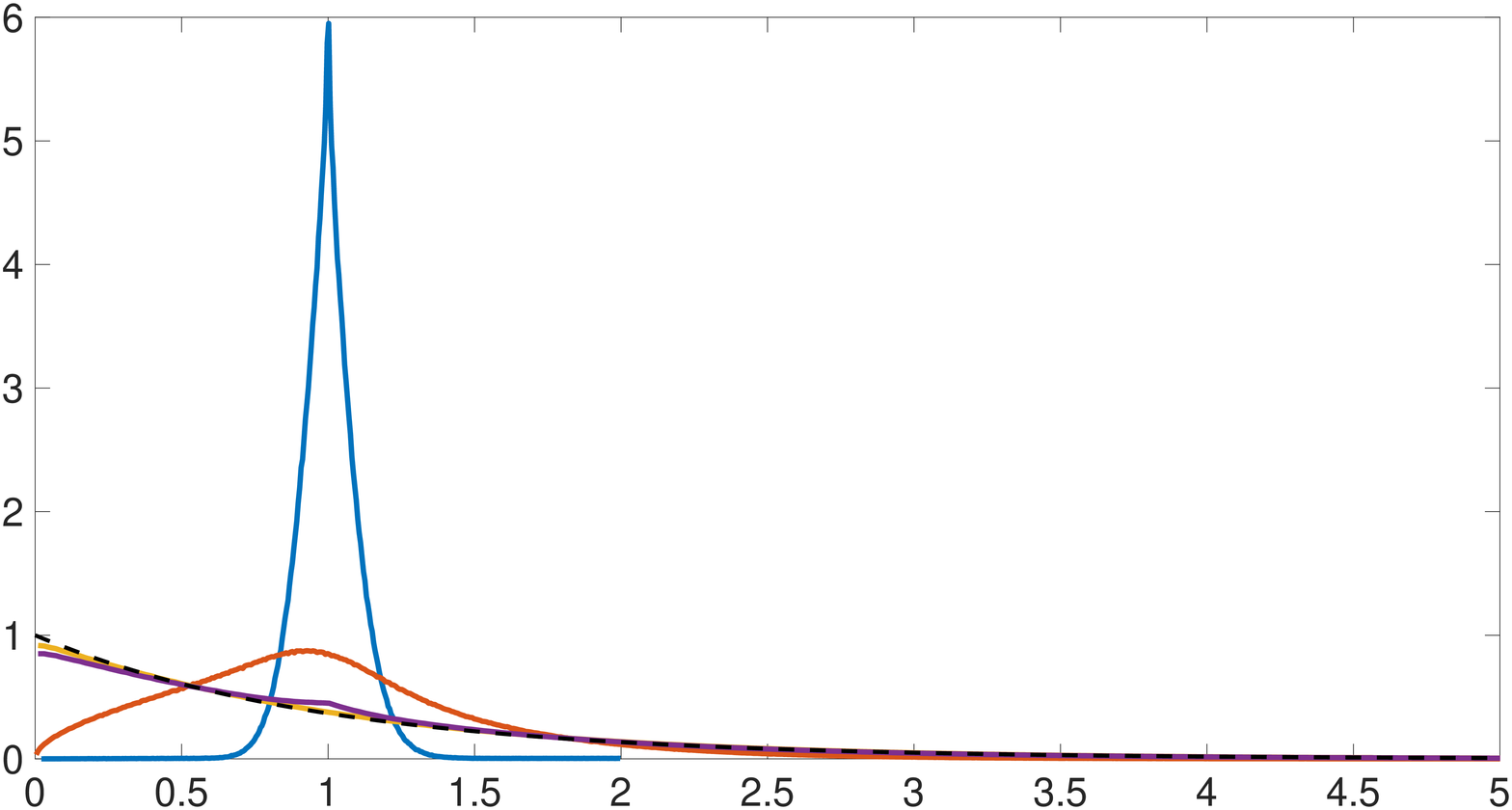}
	\caption{Distribution of the unfolded frequency spacings of the three types of disordered pendula of size $N=16384$: pure gravity (top), pure Coulomb (middle), mixed (bottom), computed from 19805, 17120 and 9060 samples, respectively. We have broken the frequency bandwidth into three different regions (displayed in Fig.~\ref{fig:prpend}) and all spacings from each region are displayed by histogram envelope curves: $\omega/N\le r_1$ (blue), $r_1<\omega/N\le r_2$ (red), $\omega/N>r_2$ (yellow). The global distribution is represented by the purple line, while for comparison we show the exponential (Poisson) distribution by the dashed line. The curve for each region is normalized to have weight one. The values for $r_1$ and $r_2$ are 0.044 and 0.143 (top), 0.032 and 0.223 (middle), 0.016 and 0.195 (bottom).}\label{fig:pspend}
\end{figure}
We have also studied numerically the normalized participation ratios (defined in Eq.~\eqref{eq:PR}) and eigenfrequency-spacing distributions of our disordered pendula. The results are presented in Figs. \ref{fig:prpend} and \ref{fig:pspend}, which exhibit similar qualitative behavior of the three pendulum types with gravitational, Coulomb or mixed interactions.

In Fig.~\ref{fig:prpend} we have plotted the participation ratio as a function of $\omega/N$. It appears that there are different scaling regimes. At higher frequencies, in the tails, $\omega/N$ is the correct scaling and we see that the normalized participation ratio decreases to zero like $\mathcal{O}(1/N)$ as $N$ increases, indicating well localized amplitude eigenvectors. As $N$ increases, the main peaks in the plots move to the left while decreasing in height at a lower rate. The exact large-$N$ scaling describing this phenomenon remains an open problem as well as the behavior very close to the origin, where we observed large participation ratio fluctuations.

Fig.~\ref{fig:pspend} displays the eigenfrequency spacing distributions for pendula of size $N=16384$ (the yellow histograms in Fig.~\ref{fig:prpend}). For the eigenfrequencies $\omega_k$, $k=1,\ldots,N$, namely the square roots of eigenvalues of each sample, we define the unfolded spacings in the following way:
\begin{equation}\label{eq:unfolding}
s_k=(\omega_{k+1}-\omega_k) \overline{\tilde\rho}(\omega_k,\omega_{k+1}), \qquad k=1,\ldots,N-1,
\end{equation}
where $\overline{\tilde\rho}(\omega_k,\omega_{k+1})$ denotes the mean of the density \eqref{mode-density} between frequencies $\omega_k$ and $\omega_{k+1}$, averaged over all samples. We have collected all spacings $s_k$ from all samples and displayed the spacing distribution by the purple histogram envelope curves in Fig.~\ref{fig:pspend}. More detailed information can be gleaned from studying the spacing distribution in various frequency subregions of the total band, which are suggested by the participation ratio curves. To this end, we have divided the frequency band into three regions: The full-width-half-maximum (FWHM) of the main peaks of the yellow histograms in Fig.~\ref{fig:prpend} (marked by the two vertical dashed lines), the high frequency range to the right of the FWHM, and the low frequency range to its left. In the right, high frequency part of the band, represented by the yellow curves in Fig.~\ref{fig:pspend}, we see that the spacing distribution is close to the exponential (Poissonian) distribution displayed by the dashed line, showing that there is no repulsion between eigenvalues. This is consistent with the localized nature of the corresponding eigenvectors, as indicated by the $\mathcal{O}(1/N)$ magnitude of their participation ratios. On the other hand, in the left, low frequency part of the band, displayed by the blue curves, the spacing distribution goes to zero for small spacings, demonstrating manifest repulsion. As we move from low frequencies into the FWHM part of the band, represented by the red curves in Fig.~\ref{fig:pspend}, we see that eigenmode repulsion decreases, despite the seizable participation ratios of the associated eigenvectors. The resolution of our numerical analysis is not fine enough to determine whether the red curves go all the way down to zero at zero spacing, indicating the absence of repulsion (although repulsion is clearly reduced the most for purely gravitational pendula). 

There is clearly a crossover between maximal repulsion at small frequencies to zero repulsion, Poissonian spacing distribution at high frequecies, but the transitionary region includes extended eigenstates with appreciable participation ratios. This is to be contrasted with the analogous crossover in the one-dimensional {\em finite size} Anderson model, with clear cut correlation between minimally localized states in the middle of the band with strong level repulsion, and maximally localized states at the band edges, with Poissonian spacing distribution. 
After all, the matrix pencil spectral problem \eqref{eigen} for our pendula, with its full and nontrivial metric $\M{M}$, high connectivity and long-range interactions, is different from the corresponding spectral problem for disordered crystals with nearest-neighbor interactions, which are avatars of Anderson's model. 

\subsection{Random Matrix Model}\label{MK}

The matrices $\M{M}$ and $\M{K}$ corresponding to small oscillations of the pendula discussed in Section \ref{pendulum} are full and rather complicated functions of the $3N$ parameters entering the problem. Such highly connected systems clearly lend themselves to analysis in terms of random matrices. 

The authors of  \cite{SKS}  have applied RMT to studying heat transfer by a highly connected and disordered network of oscillators. A considerable simplification occurring in \cite{SKS}, as compared to the systems discussed in the present paper, is that the mass matrix $\M{M}$ is simply proportional to the unit matrix. Thus, these authors needed only to apply standard RMT techniques to analyze the random matrix $\M{K}$. 

At the next level of complexity lies the analysis carried in \cite{Fyodorov} of the spectral statistics of real symmetric random matrix pencils with a deterministic diagonal metric, with nice application to fully connected electrical $LC$-networks. (See also \cite{Marchenko-Pastur, Pastur} for earlier work on such pencils.)

For an interesting recent application of RMT to studying the vibrational spectra of glassy media, in which the Marchenko-Pastur distribution plays an important role, see \cite{Zaccone}.

The methods tailored for disordered chains or crystals \cite{Dyson, Schmidt, Mattis}, as well as the more standard RMT methods used in \cite{SKS},  are inapplicable for determining the average phonon (or vibrational) spectra of systems described by full non-commuting random matrices $\M{M}$ and $\M{K}$. This requires a different approach: 

Since there is no reason to expect any statistical correlation between these two matrices, we shall draw them from two independent probability ensembles. The matrix elements of either $\M{M}$ or $\M{K}$ cannot be distributed independently. The elements of each matrix are correlated by the fact that these matrices are positive. By definition, these matrices are also real. The least biased way to fulfil these constraints is to take these matrices to be of Wishart form $\M{C}^T\M{C}$, with $\M{C}$ an $N\times N$ real Ginibre matrix\cite{Ginibre}. We shall however henceforth relax the constraint that $\M{M}$ and $\M{K}$ be real, and take them to be positive hermitian, with $\M{C}$ drawn from the complex Ginibre probability ensemble
\begin{equation}\label{ginibre}
P_\sigma(\M{C})  = \frac{1}{\cal Z} e^{-{N\over\sigma^2}{\rm Tr}\,\M{C}^\dagger\M{C}}\,,
\end{equation}
with the variance tuned such that the eigenvalues of $\M{C}$ are spread uniformly in a disk of finite radius $\sigma$ in the complex plane, as $N$ tends to infinity. Here ${\cal Z}$ is a normalization factor, and expectation values are given by 
\begin{equation}\label{expectation}
\langle F(\M{C},\M{C}^\dagger)\rangle_\sigma  = \int\left[\prod_{i,j=1}^N\!\! d\Re C_{ij}\, d\Im C_{ij}\right]\, P_\sigma(\M{C}) F(\M{C},\M{C}^\dagger)
\end{equation}
Thus, we form two such independent complex Ginibre ensembles, one for $\M{K} = \M{C}_1^\dagger\M{C}_1$ with variance $\sigma_K^2$, and another for 
$\M{M} = \M{C}_2^\dagger\M{C}_2 + m_0 $ with variance $\sigma_M^2$. The {\em positive} shift parameter $m_0$ ensures that $\M{M}\geq m_0$ is strictly positive with probability one, as it should be. 

\begin{figure}[b]
\includegraphics[width=\columnwidth]{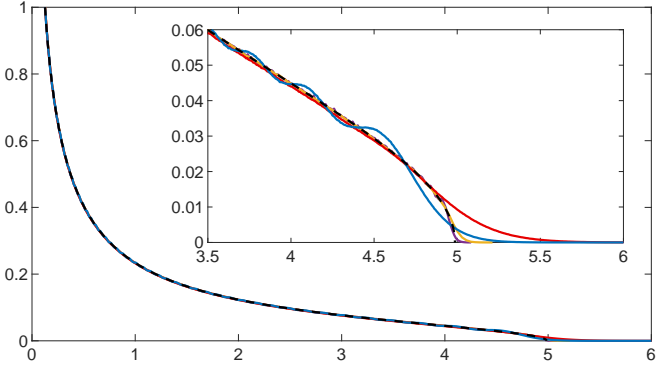}
\caption{Histogram envelope curve showing the density of eigenvalues $\varrho_{\M{H}}(\omega^2)$ of the mechanical system for complex matrices with $N=64$ (blue) and $N=1024$ (purple) and for real matrices with $N=64$ (red) and $N=1024$ (yellow), calculated from $2\times 10^7$ samples ($N=64$) and $50000$ samples ($N=1024$). The parameters of the system are $\mu=m_0=0.5$, $\sigma_M=\sigma_K=1$ (that is, $\omega_0^2=1$).  The dashed black line shows the theoretical large $N$ prediction for complex matrices given by \eqref{eq:rho}. The inset here (and in all figures below) shows a magnification of the edge behavior for the same data.}\label{fig:real_vs_complex}
\end{figure}

Such a generalization from real into complex matrices should not change the vibration spectrum in the thermodynamic limit. We have verified this expectation numerically: The difference between real and hermitian matrices amounts only to small finite-$N$ corrections at the high frequency band-edge, which vanish as $N$ tends to infinity (see Fig.~\ref{fig:real_vs_complex}). 
Of course, a more detailed investigation of the spectral statistics of such systems, such as studying level-spacings, will depend on whether one is studying real or complex matrices. In this paper we focus exclusively on the average spectrum in the thermodynamic limit, which is not affected by taking $\M{M}$ and $\M{K}$ to be complex hermitian. Issues of more detailed spectral statistics of such systems is an open problem. 

By making this innocuous generalization to complex Ginibre matrices, we can straightforwardly bring techniques of free probability theory \cite{free, burda} to bear, and use them to obtain the average spectrum analytically. 

Having defined the probability ensembles for $\M{M}$ and $\M{K}$, we draw a pair of such matrices from their corresponding ensembles and compute the random matrix $\M{H} = \M{M}^{-1}\M{K}$, in accordance with \eqref{H}. 

Our objective is to calculate the resolvent 
\begin{equation}\label{resolventH}
G_{\M{H}}(z;m_0,\sigma_M,\sigma_K) = \Big\langle {1\over N}{\rm Tr} {1\over z-\M{M}^{-1}\M{K}}\Big\rangle_{\sigma_M,\sigma_K}
\end{equation}
of $\M{H} = \left (\M{C}_2^\dagger\M{C}_2 + m_0\right)^{-1} \M{C}_1^\dagger\M{C}_1$, averaged over  $P_{\sigma_K}(\M{C}_1) $ and $P_{\sigma_M}(\M{C}_2) $ in \eqref{ginibre}, according to \eqref{expectation}, in the large-$N$ limit. We can then obtain the desired averaged density of eigenvalues 
\begin{equation}\label{Hdensity}
\varrho_{\M{H}}(\omega^2) =  \Big\langle {1\over N}{\rm Tr}\, \delta\left(\omega^2-\M{H}\right)\Big\rangle_{\sigma_M,\sigma_K}
\end{equation}
of $\M{H}$ from (\ref{resolventH}) in the usual manner \cite{QFTNut}
\begin{equation}\label{phonon-density}
\varrho_{\M{H}}(\omega^2)  = \frac{1}{\pi}{\rm Im}G_{\M{H}}(\omega^2-i\epsilon;m_0,\sigma_M,\sigma_K)
\end{equation}
as $\epsilon\rightarrow 0^+$.

An immediate consequence of \eqref{ginibre}-\eqref{resolventH} is that the resolvent \eqref{resolventH} obeys the scaling law
\begin{equation}\label{scaling}
G_{\M{H}}(z;m_0,\sigma_M,\sigma_K) = \left(\frac{\sigma_M}{\sigma_K}\right)^2 G_{\M{H}}\left(\zeta; \mu,1,1\right)\,,
\end{equation}
with rescaled variables
\begin{equation}\label{scaling-variables}
\zeta = \left(\frac{\sigma_M}{\sigma_K}\right)^2 z = \frac{z}{\omega_0^2}\quad {\rm and}\quad \mu = \frac{m_0}{\sigma_M^2}\,.
\end{equation}
Clearly, $\sigma_M^2$ has dimensions of mass, and $\sigma_K^2$ has dimensions of force per unit length. Thus, $\frac{\sigma_K^2}{\sigma_M^2} = \omega_0^2$ has dimensions of frequency squared, and \eqref{scaling-variables} simply instructs us to measure $m_0$ in units of $\sigma_M^2$ and the complex spectral parameter $z$ in units of $\omega_0^2$.  

For later use, let us also introduce the density of eigenvalues associated with $G_{\M{H}}\left(\zeta; \mu,1,1\right)$, namely, 
\begin{equation}\label{phonon-density1}
\varrho_{\M{H}}(x;\mu)  = \frac{1}{\pi}{\rm Im}G_{\M{H}}(x-i\epsilon;\mu,1,1)
\end{equation}
with $x={\rm Re}\zeta$. It then follows from \eqref{phonon-density} -\eqref{scaling-variables} (or directly from \eqref{Hdensity} upon substituting $x=(\omega/\omega_0)^2)$) that 
\begin{equation}\label{density-relation}
\varrho_{\M{H}}(\omega^2) = {1\over\omega_0^2}\varrho_{\M{H}}\left({\omega^2\over\omega_0^2};\mu\right).
\end{equation}

\subsection{Free Probability Theory}\label{free} 
The random matrix $\M{H}$ is the product of two statistically independent, positive-definite random matrices, taken from unitary-invariant probability ensembles. Indeed, the random matrix $\M{X} = \M{M}^{-1}$, like $\M{M}$, has a probability distribution invariant under unitary rotations. 
After a straightforward calculation, one obtains its probability distribution as 
\begin{equation}\label{X}
Q(\M{X})  = {\cal N} {\Theta\left(\M{X}^{-1} -m_0\right)\over \left(\det \M{X}\right)^{2N}} e^{-{N\over \sigma_M^2}{\rm Tr}\,(\M{X}^{-1} - m_0)}\,,
\end{equation}
where the matricial step function $\Theta(\cdot)$ enforces positivity of $\M{X}^{-1} -m_0 ~ (= \M{C}_2^\dagger\M{C}_2)$, and ${\cal N}$ is a normalization factor.

The {\em S-transform} of free probability theory \cite{free,burda} is a common tool for calculating the resolvent and density of eigenvalues of products like $\M{H}$, in the large-$N$ limit. In our case, it reduces the calculation of \eqref{resolventH} to solving a certain cubic equation (see \eqref{cubic}). The procedure is as follows: Compute the resolvents of $\M{X}$ and $\M{K}$, then compute the $S$-transfroms of these resolvents and multiply them together to obtain the $S$-transform of $\M{H}$, and finally, make an inverse transform of the latter to obtain the resolvent \eqref{resolventH} of $\M{H}$.

The resolvent 
\begin{equation}\label{resolvent}
G(z;\sigma) =\Big \langle {1\over N}{\rm Tr} {1\over z-\M{C}^\dagger\M{C}}\Big\rangle_{\sigma} = \frac{1}{2\sigma^2}\left(1-\sqrt{1-\frac{4\sigma^2}{z}}\right)
\end{equation}
is a special case of a more general expression obtained long ago by Marchenko and Pastur \cite{Marchenko-Pastur}, and it can be derived in several ways \cite{rectangles}. It is analytic in the cut complex-$z$ plane, with branch points at $z=0,4\sigma^2$.The cut emanating from each branch point runs to the left, along the real axis. With this assignment of the cuts, $G(z;\sigma) $ is pure-imaginary along the segment $[0,4\sigma^2]$, which is the support of the average density of eigenvalues 
\begin{equation}\label{density}
\rho(x;\sigma)  = \frac{1}{\pi}{\rm Im}G(x-i\epsilon;\sigma)  = \frac{1}{2\pi\sigma^2}\sqrt{\frac{4\sigma^2-x}{x}}\,.
\end{equation}
From the large-$z$ expansion $G(z;\sigma) = \frac{1}{z} + \frac{\sigma^2}{z^2}+\ldots$, we can read-off the first moment  $\mu_1 = \langle \frac{1}{N}{\rm Tr} \left(\M{C}^\dagger\M{C}\right)\rangle_\sigma = \sigma^2 $, as should be expected from \eqref{ginibre}. 

The resolvent $G_{\M{K}}(z;\sigma_K) $ of $\M{K}$ is simply $G(z;\sigma_K)$, and its first moment is non-vanishing. (The latter is a technical requirement for applying the $S$-transform.) The resolvent of $\M{X} = \M{M}^{-1}$ can also be obtained from \eqref{resolvent} in a straightforward manner:
\begin{eqnarray}\label{resolventX}
&&G_{\M{M}^{-1}}(z;m_0,\sigma_M) = \Big\langle {1\over N}{\rm Tr} {1\over z-\M{M}^{-1}}\Big\rangle_{\sigma_M}\nonumber \\
&=& {1\over z} -{1\over z^2}\Big\langle {1\over N}{\rm Tr} {1\over z^{-1}-\M{M}}\Big\rangle_{\sigma_M}\nonumber\\ &=& {1\over z} -{1\over z^2}\Big\langle {1\over N}{\rm Tr} {1\over z^{-1}-m_0 -\M{C}_2^\dagger\M{C}_2}\Big\rangle_{\sigma_M}\nonumber\\&=&
{1\over z} -{1\over z^2}G(z^{-1}-m_0;\sigma_M)\nonumber\\
&=&{1\over z} -{1\over 2\sigma_M^2z^2} + {1\over 2\sigma_M^2z^2}\sqrt{1-(m_0+ 4\sigma_M^2)z\over 1-m_0z}\,.
\end{eqnarray}
The cut structure for this resolvent is similar to that of \eqref{resolvent}, with the average density of eigenvalues 
\begin{eqnarray}\label{densityX}
\varrho_{\M{M}^{-1}}(x)  &=& \frac{1}{\pi}{\rm Im}G_{\M{M}^{-1}}(x-i\epsilon;m_0,\sigma_M) \nonumber\\ &=& 
\frac{\left(1 + {4\sigma_M^2\over m_0}\right)^{1\over 2}}{2\pi\sigma_M^2}{1\over x^2}\sqrt{\frac{x-{1\over m_0+ 4\sigma_M^2}}{{1\over m_0}-x}}
\end{eqnarray}
supported between the two branch points along the segment $[(m_0+ 4\sigma_M^2)^{-1}, m_0^{-1}]$\,. The large-$z$ expansion of $G_{\M{M}^{-1}}(z;m_0,\sigma_M)$ yields the first moment $\mu_1 = \langle \frac{1}{N} {\rm Tr} \M{M}^{-1}\rangle_{\sigma_M} = {1\over 2\sigma_M^2}(\sqrt{1 + {4\sigma_M^2\over m_0}}-1)>0 $\,.

The next step \cite{burda} in computing the $S$-transform of any of the aforementioned resolvents $G(z)$ amounts to defining a related function 
\begin{equation}\label{phi}
\phi(z) = {1\over z} G\left({1\over z}\right) -1\,.
\end{equation}
Recall that  $G(z)$ is the generating function for the moments $\mu_n = \langle \frac{1}{N}{\rm Tr} \M{A}^n\rangle$ of the positive definite random matrix $\M{A} (=\M{K}, \M{M}^{-1})$:
\begin{equation}\label{generating}
G(z) = \sum_{n=0}^\infty {\mu_n\over z^{n+1}}\,,
\end{equation}
where of course $\mu_0=1$, and $\mu_1>0$ by assumption. Thus, 
\begin{equation}\label{generating1}
\phi(z) = \sum_{n=1}^\infty \mu_n z^n\,.
\end{equation}  
In particular, this means that 
\begin{equation}\label{condition}
\lim _{z\rightarrow 0} {\phi(z)\over z} = \mu_1>0\,.
\end{equation}
Let $\chi (u)$ be the functional inverse of $\phi(z)$, that is the solution of 
\begin{equation}\label{inverse}
\chi(\phi(z))=z\,, 
\end{equation}
consistent with (\ref{condition}). Thus, in case \eqref{inverse} has several roots for $\chi(u)$, we pick that root which behaves like $\frac{u}{\mu_1}$ as $u\rightarrow 0$.  Then, finally, the $S$-transform of $G(z)$ is defined as 
\begin{equation}\label{S}
S(u) = {u+1\over u} \chi (u)\,.
\end{equation}
Following this procedure, we thus obtain the $S$-transforms of $G_{\M{K}}(z;\sigma_K)$ and $G_{\M{M}^{-1}}(z;m_0,\sigma_M),$ respectively, as
\begin{eqnarray}\label{S-transforms}
S_{\M{K}}(u) &=&\!\! {1\over \sigma_K^2 (u+1)}\nonumber\\
S_{\M{M}^{-1}}(u) &=&\!\! {\sigma_M^2\over 2}\!\!\left[{m_0\over \sigma_M^2}\! - u +\sqrt{\left(u+{m_0\over\sigma_M^2}\right)^2 \!\!\!+ \!\!{4m_0\over\sigma_M^2}}\right]
\end{eqnarray}
We now multiply the two expressions in \eqref{S-transforms} to obtain the $S$-transform of $\M{H}$, 
\begin{equation}\label{SH}
S_{\M{H}}(u) = S_{\M{M}^{-1}}(u)S_{\M{K}}(u) = {u+1\over u}\chi_{\M{H}}(u)\,.
\end{equation}
We thus find
\begin{equation}\label{chiH}
\chi_{\M{H}}(u) = {\sigma_M^2\over 2\sigma_K^2}{u\over (u+1)^2}\!\!\left[{m_0\over \sigma_M^2}\! - u +\sqrt{\left(u+{m_0\over\sigma_M^2}\right)^2 \!\!\!+ \!\!{4m_0\over\sigma_M^2}}\right],
\end{equation}
which is the functional inverse of $\phi_{\M{H}}(z)$. After some work, we thus obtain from (\ref{inverse}) a quartic equation for $\phi_{\M{H}}(z)$, which contains a factor 
$\phi_{\M{H}} +1$. Since  $\phi_{\M{H}}$ is not identically constant, we can safely divide through by this factor and obtain a cubic equation for  $\phi_{\M{H}}$. Finally, by virtue of \eqref{phi}, this cubic equation leads to a cubic equation for $G_{\M{H}}(z;m_0,\sigma_M,\sigma_K)$:
\begin{eqnarray}\label{cubic}
&&\left[\left({\sigma_K\over\sigma_M}\right)^4 z + \left({\sigma_K\over\sigma_M}\right)^2 z^2 \right] G_{\M{H}}^3 \nonumber\\
&-& \left[\left(2+{m_0\over\sigma_M^2}\right)\left({\sigma_K\over\sigma_M}\right)^2 z + {m_0\over \sigma_M^2}z^2\right] G_{\M{H}}^2\nonumber\\ 
&+& \left[\left(1+{m_0\over\sigma_M^2}\right)\left({\sigma_K\over\sigma_M}\right)^2  + 2{m_0\over \sigma_M^2}z\right] G_{\M{H}} - {m_0\over\sigma_M^2} = 0\,.\nonumber\\
\end{eqnarray}
The desired resolvent \eqref{resolventH} is that root of \eqref{cubic} with asymptotic behavior $G_{\M{H}}(z)\sim \frac{1}{z}$ as $z\rightarrow\infty$. This equation is consistent with the scaling law \eqref{scaling}, and we can write it in terms of the rescaled variables \eqref{scaling-variables} more compactly as 
\begin{eqnarray}\label{scaled-cubic}
\left(\zeta + \zeta^2\right) \Gamma(\zeta,\mu)^3 -\left[(2+\mu)\zeta + \mu\zeta^2\right]\Gamma(\zeta,\mu)^2 \nonumber\\
+ (1+\mu+2\mu\zeta)\Gamma(\zeta,\mu) - \mu=0\,,
\end{eqnarray}
where we have defined 
\begin{equation}\label{Gamma}
\Gamma(\zeta,\mu) = \left({\sigma_K\over\sigma_M}\right)^2 G_{\M{H}}(z;m_0,\sigma_M,\sigma_K) = G_{\M{H}}(\zeta;\mu,1,1) 
\end{equation}
\begin{figure}[b]
\includegraphics[width=\columnwidth]{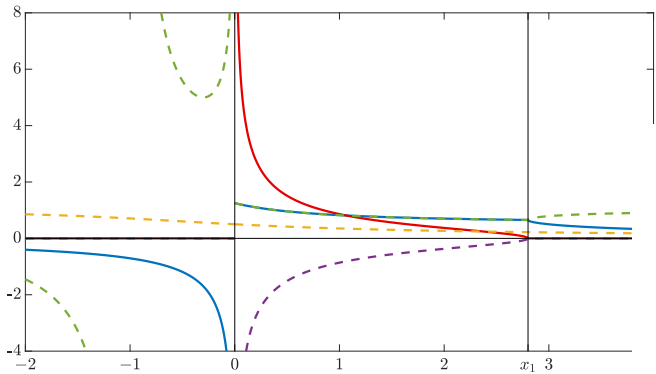}
\caption{The solutions $\Gamma$ of the cubic equation \eqref{scaled-cubic} as a function of $\zeta$, along the real $\zeta-$axis for $\mu=1$. The color code for the three solutions is as follows: The real (imaginary) part of solution 1 is blue (red), the real (imaginary) part of solution 2 is green (purple), and the purely real solution 3 is yellow. The nonphysical solutions 2 and 3 are plotted in dashed lines. Note that solutions 1 and 2 are complex-conjugate in the interval $(0,x_1)$, and that their imaginary parts vanish identically outside this interval.}\label{fig:solutions}
\end{figure}
The solutions of the cubic equation \eqref{scaled-cubic} are illustrated in Fig.~\ref{fig:solutions}. 

\subsection{Analytical Derivation of the Density of Eigenvalues}\label{density of modes} 
\begin{figure}[b]
\begin{center}
\includegraphics[width=\columnwidth]{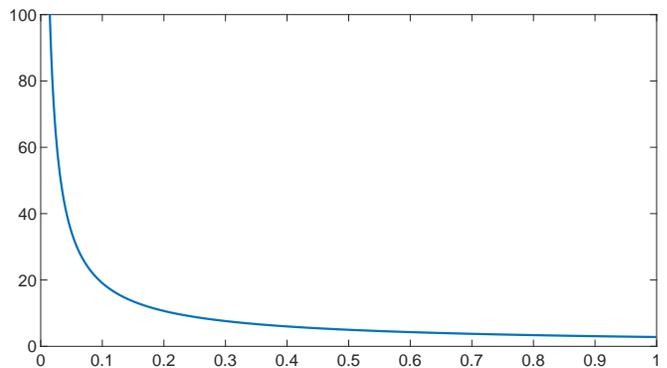}
\caption{The density of eigenvalues $\varrho_{\M{H}}(\omega^2)$ is supported along the interval $0\leq\omega^2\leq \omega_0^2 x_1$. The graph shows the rescaled right endpoint $x_1$, given by \eqref{eq:x1}, as a function of $\mu$.}\label{fig:endp}
\end{center}
\end{figure}

Our main objective is to calculate $\varrho_{\M{H}}(\omega^2)$. Based on \eqref{density-relation} it is enough to determine $\varrho_{\M{H}}(x;\mu)=\frac{1}{\pi}\Im \Gamma (x-i\epsilon;\mu)$. Therefore we should look for complex solutions of \eqref{scaled-cubic} along the $\Re\zeta=x$ axis.
More specifically, we are looking there for a pair of complex-conjugate roots. Thus, we should analyze the discriminant 
\begin{equation}
\Delta_\Gamma(x)=x p_3(x), \label{eq:DeltaGamma}
\end{equation}
of \eqref{scaled-cubic} and find where it is negative. In \eqref{eq:DeltaGamma} $p_3$ is the cubic polynomial
\begin{align*}
p_3&(x)=(\mu+4)\mu^3 x^3 +2\mu^2(\mu^2+2\mu-6) x^2 \nonumber \\
&+(\mu^3-4 \mu^2-20 \mu+12)\mu x -4(\mu^3+3\mu^2+3\mu+1).  
\end{align*} 
The discriminant of $p_3$ is 
\begin{equation*}
\Delta_{p_3}=-16\mu^8 (\mu^2+10\mu+27)^3 \label{eq:Deltap3}
\end{equation*}
which is manifestly negative for all $\mu>0$, meaning $p_3(x)$ has only one real root
\begin{align}
x_1=&\frac{1}{3\mu^3(\mu+4)}\Big(-2\mu^2(\mu^2+2\mu-6) \nonumber\\
&\qquad+\frac{(\xi_1+\sqrt{\Delta_1})^{1/3} +(\xi_1-\sqrt{\Delta_1})^{1/3}}{2^{1/3}} \Big), \label{eq:x1}
\end{align}
 with
\begin{align*}
\xi_1&=2\mu^7(\mu^5+24\mu^4+264\mu^3+1574\mu^2+4806\mu+5832),\\
\Delta_1&=432\mu^{14}(\mu+4)^2 (\mu^2+10\mu+27)^3.
\end{align*}
It can be shown that $x_1$ is positive. Thus for positive $\mu$, $\Delta_\Gamma(x)$ is negative {\em only} along the interval $(0,x_1)$, and this is where \eqref{scaled-cubic} has a pair of complex-conjugate roots. This interval, or more precisely  $(0,\omega_0^2 x_1)$, is therefore the desired support of $\varrho_{\M{H}}(\omega^2)$, as can be inferred from \eqref{scaling}-\eqref{scaling-variables}. This support is purely positive as it should be, due to positivity of the matrix $\M{H}$. The endpoint $x_1$ as a function of $\mu$ is plotted in Fig.~\ref{fig:endp}. We see that $x_1(\mu)$ is a monotonically decreasing function, which should be expected physically because $\omega^2\sim \M{M}^{-1}\M{K}\sim 1/\mu$. 

Picking that root of the cubic \eqref{scaled-cubic} which has positive imaginary part along $[0,x_1]$ we thus find 
\begin{equation}
\varrho_{\M{H}}(x;\mu)=\frac{1}{2\sqrt{3} x(x+1)\pi}\left(\frac{(\xi_\Gamma+\delta_\Gamma)^{1/3}}{2^{1/3}}-\frac{2^{1/3}\chi_\Gamma}{(\xi_\Gamma+\delta_\Gamma)^{1/3}}\right),\label{eq:rho}
\end{equation}
where
\begin{align*}
\xi_\Gamma&=%\xi_\Gamma(x)=
-x^2 [2\mu^3 x^4 +6\mu^2(\mu-1) x^3 +3\mu(2\mu^2-7\mu+2) x^2 \nonumber \\ %\label{eq:XiGamma}\\ 
&\qquad\qquad\quad+2(\mu^3-12\mu^2+3\mu-1)x -9(\mu^2+2)], \\
\delta_\Gamma&=%\delta_\Gamma(x)=
x(x+1)\sqrt{-27\Delta_\Gamma(x)}, \\ %\label{eq:deltaGamma}
\chi_\Gamma&=%\chi_\Gamma(x)=
\mu^2 x^4+2\mu(\mu-1)x^3 +(\mu^2\!-5\mu+1)x^2-3(\mu+1)x. %\label{eq:chiGamma}
\end{align*}
Thus, by substituting $x=(\omega/\omega_0)^2$ in \eqref{eq:rho} and using \eqref{density-relation}, we obtain $\varrho_{\M{H}}(\omega^2)$ as desired.

$\varrho_{\M{H}}(x;\mu)$ diverges at the origin like $1/\sqrt{x}$, and vanishes at $x_1$ like $\sqrt{x_1-x}$ (with known coefficients). This divergence of $\varrho_{\M{H}}(x;\mu)$ like $1/\sqrt{x}\sim 1/\omega$ at the origin is reminiscent the behavior \eqref{universal} of pendula, indicating a {\em universal} such behavior of the density of vibration eigenmodes of highly connected systems at low frequencies. 

\begin{figure}[b]
\includegraphics[width=\columnwidth]{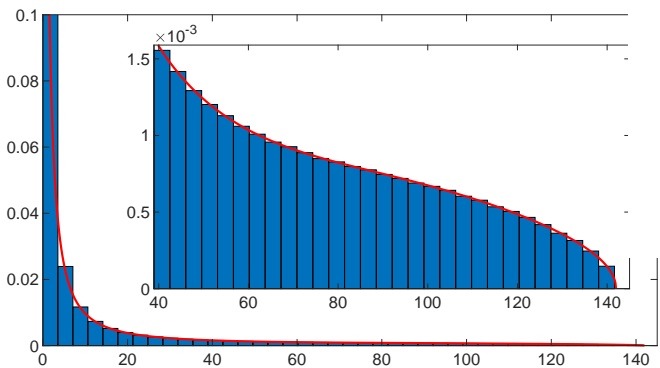}
\caption{Histogram of a simulation of $\varrho_{\M{H},N}^\text{num}(\omega^2)$ using one sample with $N=
65536$, $\mu = 0.01, \sigma_M = \sigma_K = 1$ ($\omega_0^2=1$). The red line
shows the theoretical large-$N$ prediction given by \eqref{eq:rho}.}\label{Fig:hist}
\end{figure}
\begin{figure}[b]
\includegraphics[width=\columnwidth]{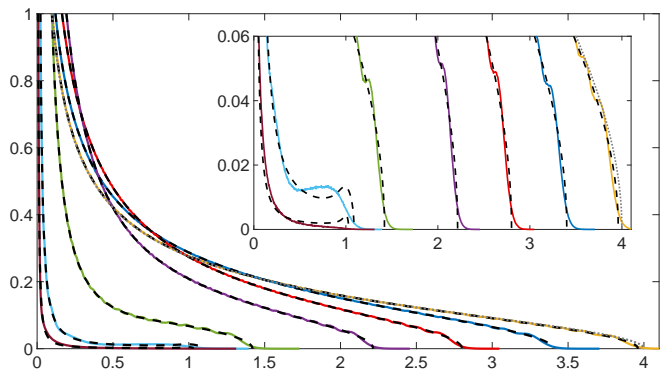}
\caption{Plot of the density of eigenvalues $\varrho_{\M{H},N}^\text{num}(\omega^2)$ from a simulation using $10^6$ samples for the fixed parameters $N=128$, $\sigma_K=1$, and $m_0=1$, for various values of $\sigma_M$, in order of increasing endpoints $\omega_0^2 x_1 = (\sigma_K/\sigma_M)^2x_1$: 500 (brown), 100 (cyan), 10 (green), 2 (purple), 1 (red), 0.5 (blue) and 0.1 (yellow). The corresponding dashed black lines show the theoretical large-$N$ prediction given by \eqref{eq:rho}. The dotted line (close to the yellow line) shows the Marchenko-Pastur distribution.}\label{Fig:sigma}
\end{figure}
\begin{figure}[b]
\includegraphics[width=\columnwidth]{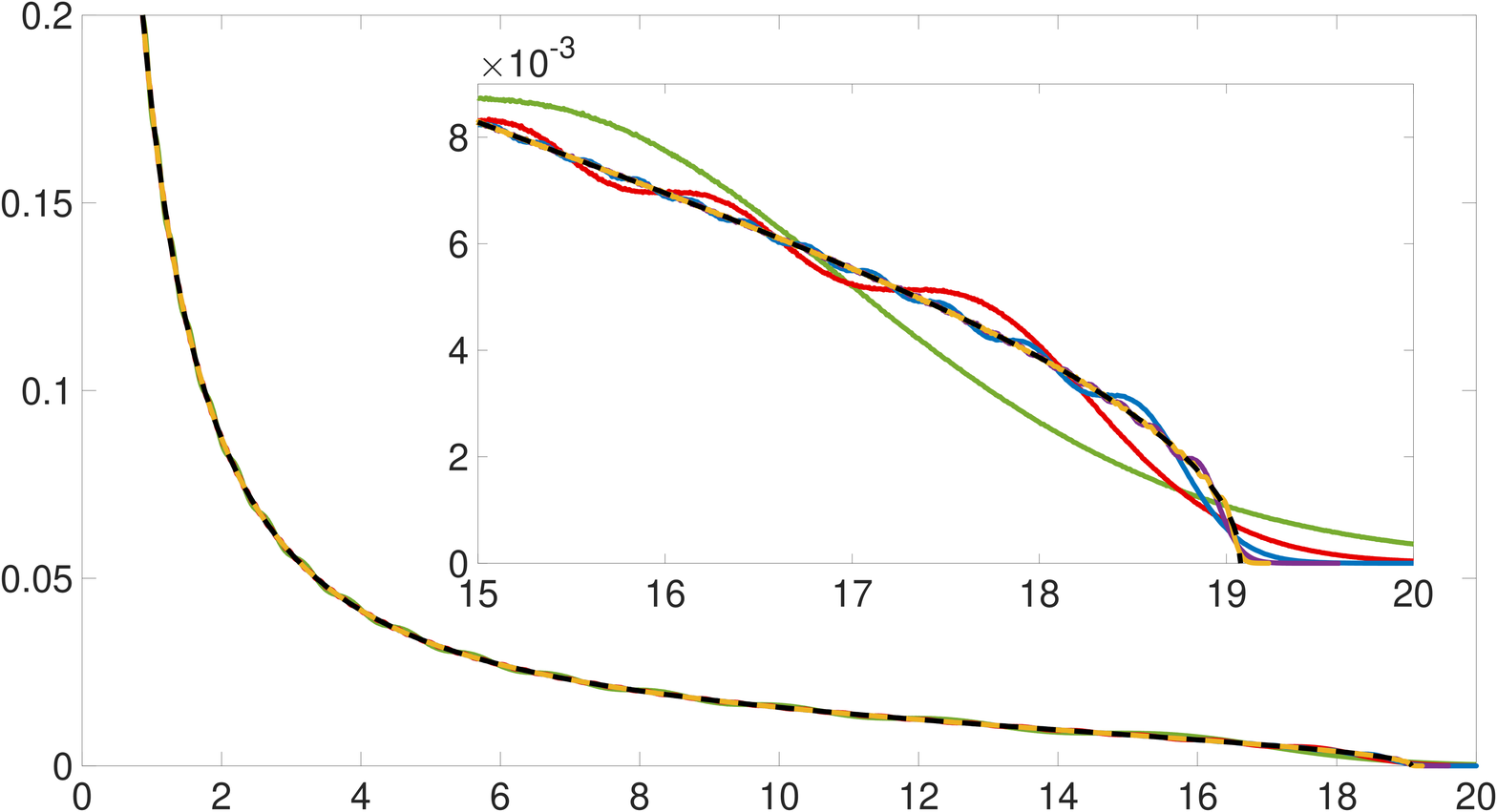}
\caption{Curves from histograms of the density of eigenvalues $\varrho_{\M{H},N}^\text{num}(\omega^2)$ from simulation using millions of samples with $N=32$ (green), $N=128$ (red), $N=512$ (blue), $N=2048$ (purple) and $N=8192$ (yellow). The parameters of the complex model are $\mu=0.1$, $\sigma_M=\sigma_K=1$ ($\omega_0^2=1$). The dashed black line shows the theoretical large-$N$ prediction given by \eqref{eq:rho}.}\label{Fig:N}
\end{figure}

Let $\varrho_{\M{H},N}^\text{num}(\omega^2)$ denote the finite $N$ averaged density of the mechanical system from numerical simulations. It is described by the curves in the following plots. They are the envelope curves gleaned from histograms with very narrow bins. These curves for $\varrho_{\M{H},N}^\text{num}(\omega^2)$ are in excellent agreement, when $N$ is large, with the analytical expression for $\varrho_{\M{H}}(\omega^2)$ as obtained from \eqref{density-relation} and \eqref{eq:rho}. In Fig.~\ref{Fig:hist} we show a histogram from one sample which demonstrates self-averaging of the eigenvalue density of large matrices. In Fig.~\ref{Fig:sigma} we display $\varrho_{\M{H},N}^\text{num}$ obtained by averaging over a million samples, and compare it to $\varrho_{\M{H}}$, for various values of the parameter $\sigma_M$. Note that in the limit $1/\mu\rightarrow 0$, $\varrho_{\M{H},N}^\text{num}(\omega^2)$ converges to the Marchenko-Pastur density (dotted line in Fig.~\ref{Fig:sigma}) as should be expected, because in this limit the matrix $\M{M}$ becomes deterministic and proportional to the unit matrix. In the opposite limit of large $\sigma_M$ (or equivalently $\mu\rightarrow 0$), the density looks qualitatively different from the Marchenko-Pastur profile. In fact, at $\mu=0$, \eqref{scaled-cubic} reduces to a quadratic equation. 

\subsection{Universal Edge Behavior of the Density}\label{universality} 
\begin{figure}[b]
\includegraphics[width=\columnwidth]{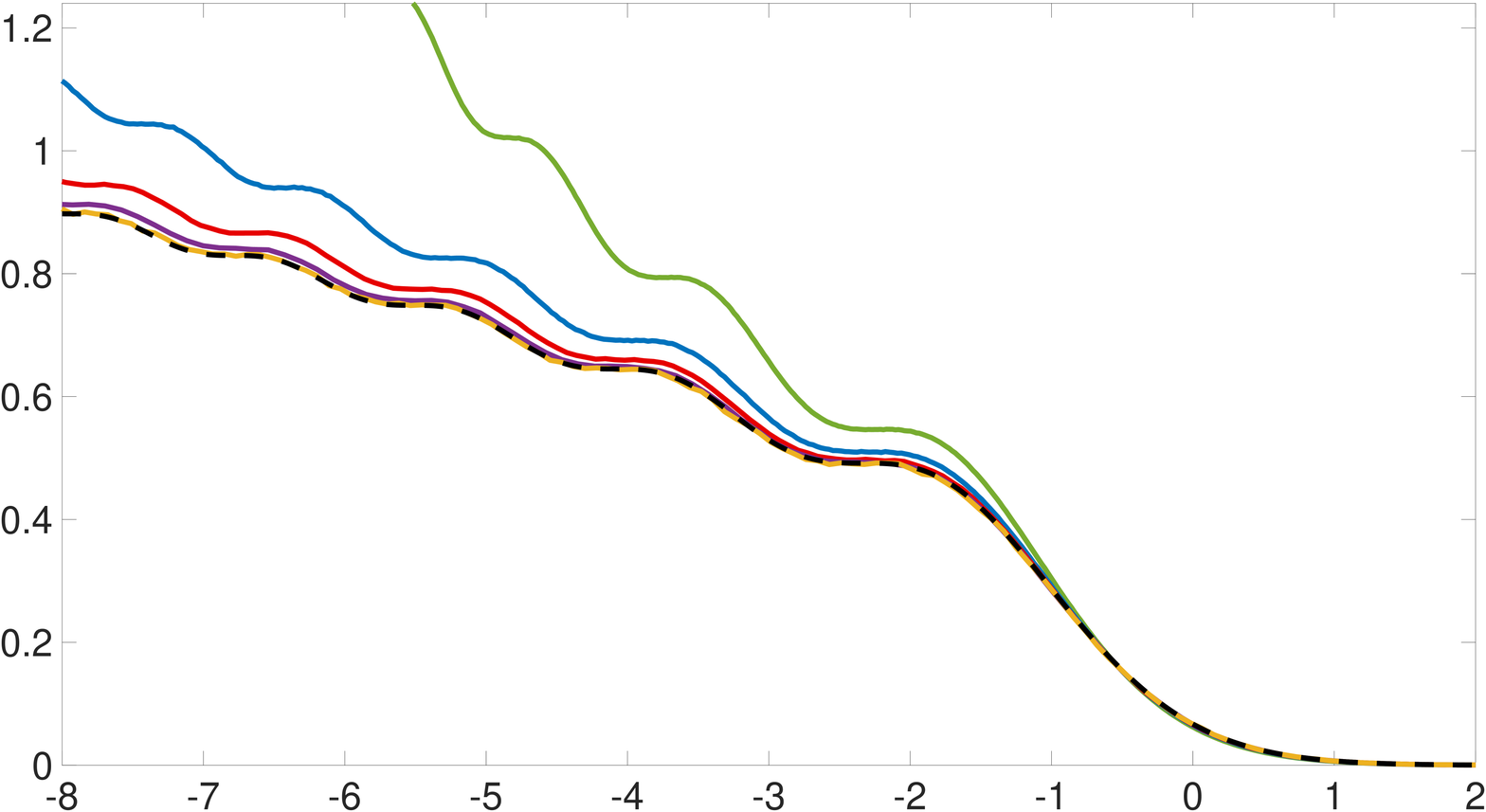}
\caption{Plot of the density of eigenvalues $\varrho_{\M{H},N}^\text{edge}(\eta)$ near the edge from a simulation of the mechanical model with complex matrices. The parameters and colors are as in Fig.~\ref{Fig:N}. 
The dashed line shows $\rho^\text{Airy}$ given in \eqref{eq:airy} and is the expected universal behavior at the edge. The parameter $r$ in \eqref{eq:rhorescaled} has been chosen as $r\approx 19.7$ to give good matching. }\label{Fig:edge}
\end{figure}

In Fig.~\ref{Fig:N} we show numerical results for the density, with a fixed choice of parameters (corresponding to having $\omega^2_0=1$ and $x=\omega^2$) and for various values of $N$. Convergence of the numerical results to the theoretical large-$N$ curve in the bulk is rapid, whereas convergence close to the high frequency (soft) edge is non-uniform, with visible finite-$N$ corrections. The model with complex matrices clearly exhibits oscillatory behavior (see  Fig.~\ref{Fig:N}) towards the high frequency edge, as in the canonical GUE case.
On the other hand, the model with real matrices has non-oscillatory edge behavior, like in the GOE case (as can be seen in Fig.~\ref{fig:real_vs_complex}).
Referring back to the complex case, we expect its edge behavior to be in the Airy universality class, because in the large-$N$ limit the density vanishes at the edge like $\sqrt{x_1-x}$. We verified this expectation numerically as can be seen in Fig.~\ref{Fig:edge}. To this end we studied the rescaled density  
\begin{equation}\label{eq:rhorescaled}
\varrho_{\M{H},N}^\text{edge}(\eta)=r N^{1/3} \varrho_{\M{H},N}^\text{num}(x_1+ r N^{-2/3} \eta),
\end{equation}
where $r=r(\mu)$ is an $N$-independent parameter. This $\varrho_{\M{H},N}^\text{edge}$ seems to converge \emph{uniformly} to the diagonal part of the Airy kernel, 
\begin{equation}
\rho^\text{Airy}(\eta)= (\mathrm{Ai}'(\eta))^2-\mathrm{Ai}''(\eta)\mathrm{Ai}(\eta)=(\mathrm{Ai}'(\eta))^2-\eta (\mathrm{Ai}(\eta))^2. \label{eq:airy}
\end{equation} 

\subsection{Eigenvectors and their Participation Ratio}\label{participation} 
\begin{figure}[b]
\includegraphics[width=\columnwidth]{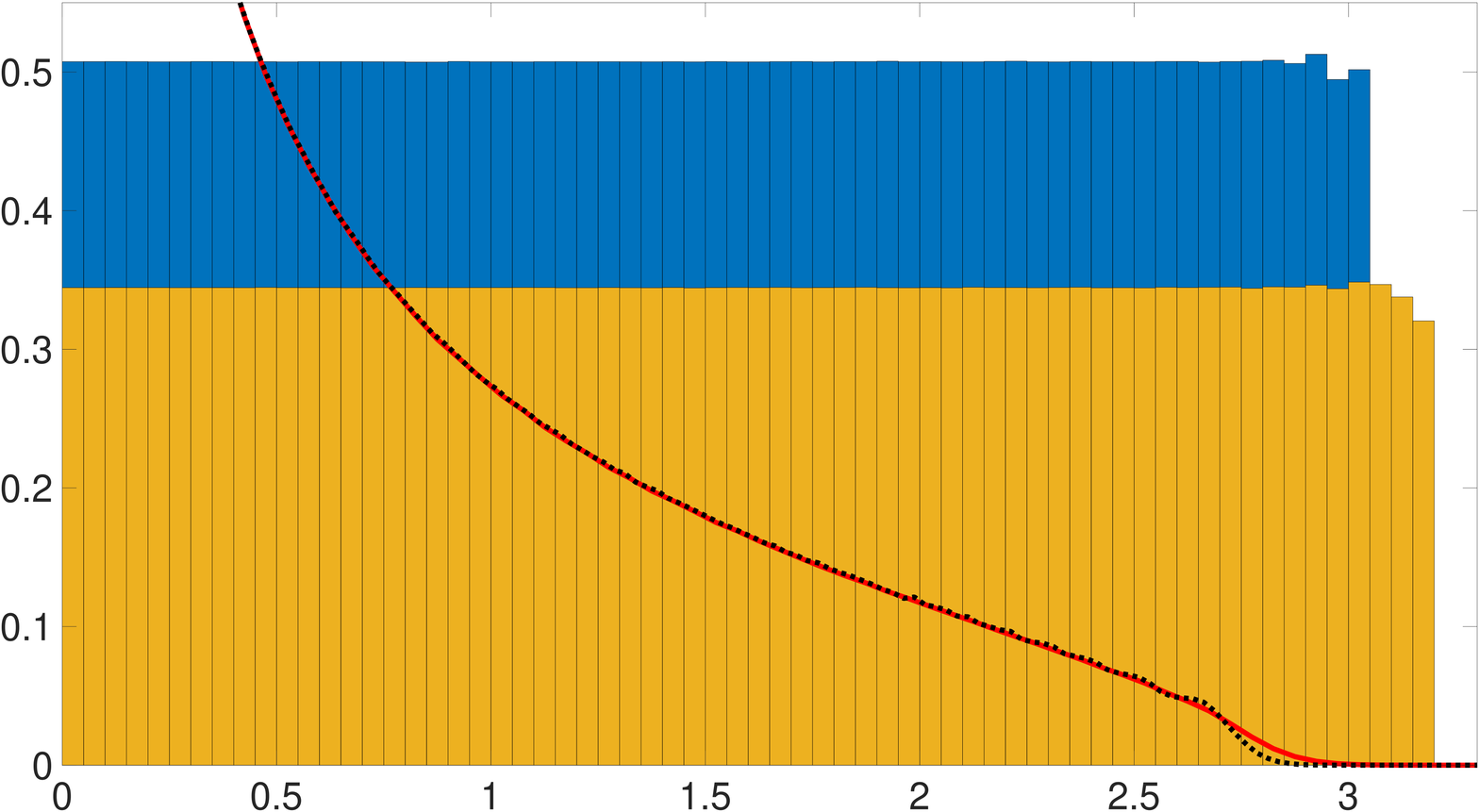}
\caption{Histogram showing the participation ratio $p(\omega^2)$ (see Eq.~\eqref{eq:PR}) of the model with parameters $N=128$, $\mu=\sigma_M=\sigma_K=1\, (\omega_0^2=1)$ for the complex (blue) and real (yellow) case using $10^5$ samples. The black dotted and red solid and line shows the density of eigenvalues for the corresponding simulations. Fluctuations in the histograms at the right edge of the spectrum are artifacts due to low statistics where the density vanishes.}\label{Fig:PR}
\end{figure}
Consider the \emph{normalized} eigenvector $\M{A}(\omega^2)$ of $\M{H}$ in \eqref{eigen}, corresponding to eigenvalue $\omega^2$ with components $A_\ell$. 
The (normalized) \emph{participation ratio}\cite{wegner} is the function of $\omega^2$
\begin{equation}\label{eq:PR}
p(\omega^2)=\frac{1}{N}\Big\langle\frac{1}{\sum_{\ell=1}^N |A_\ell(\omega^2)|^4}\Big\rangle_{\sigma_M,\sigma_K}. 
\end{equation}
It is a measure of the fraction (out of $N$) of degrees of freedom of the system that are effectively involved in a given state of vibration. In addition to the participation ratios of eigenvectors of the three types of pendula, displayed in  Fig.~\ref{fig:prpend}, we have also computed numerically the participation ratios for the complex and real matrix models. 

For random matrices, contrary to the case of pendula, our numerical results (see Fig.~\ref{Fig:PR}) show that these participation ratios are independent of the eigenvalue $\omega^2$ and converge to constants as $N$ increases. This means that in both complex and real matrix models all states are extended. That is, in all vibrational modes, essentially all degrees of freedom oscillate with amplitudes of the same order of magnitude. In other words, vibration eigenmodes tend to be collective throughout the frequency band. Moreover, these constant values seem to be universal (for various values of $\mu$) and approach (for large $N$) the value of 0.50 in the complex case and 0.33 for the real model. Interestingly, these numerical results coincide, respectively, with the participation ratios of the canonical GUE and GOE (even for low $N$). In the large $N$ limit it is straightforward to compute the GUE and GOE participation ratios from the Porter-Thomas probability distributions\cite{haake} for eigenvector components upon neglecting correlations between eigenvector components. In hindsight, this coincidence with GUE and GOE is perhaps not surprising and hints at a broader universality of our results.

\section{The Liouvillian and Diagrammatic Derivation of $S$-Transforms}\label{Liuovillian}
Derivation of the $S$-transform formula for multiplying statistically independent hermitian random matrices by means of large-$N$ planar diagrams was given in \cite{BJN}. The idea is simple and elegant: Let $\M{A}$ and $\M{B}$ be two $N\times N$ statistically independent random matrices. In order to calculate the averaged resolvent of $\M{AB}$ one should study the resolvent of the doubled-size matrix 
\begin{equation}\label{BJN}
\M{Q} = \left(\begin{array}{cc} \M{0} & \M{A}\\ \M{B} & \M{0}\end{array}\!\!\right),   
\end{equation}
because the upper diagonal block of this resolvent
\begin{eqnarray}\label{BJN-resolvent}
{1\over w-\M{Q}} &=&  \left(\begin{array}{cc} {w\over w^2 - \M{AB}} & {1\over w^2 - \M{AB} }{\small\small\small\small\small{\M{A}}} \\ {1\over w^2 - \M{BA}}{\small\small\small\small\small{\M{B}}}& {w\over w^2 - \M{BA}}\end{array}\!\!\right)
\end{eqnarray}
is essentially the desired resolvent of $\M{AB}$. By expanding the left-hand-side of \eqref{BJN-resolvent} in inverse powers of $w$ and averaging over the appropriate powers of $\M{Q}$, we obtain the diagrammatic expansion of \eqref{BJN-resolvent}. In the large-$N$ limit only planar diagrams survive, in which lines cannot cross. From this fact, and from the block structure of $\M{Q}$ and its powers, one can prove the $S$-transform product formula by consistently resumming planar diagrams.  

We should comment that a tacit assumption made by the authors of \cite{BJN} is that the spectra of $\M{AB}$ and $\M{BA}$ are real, because the perturbative expansion in powers of $w^{-1}$ assumes analyticity of all the resolvents involved off the real axis. This means that $\M{AB}$ should be {\em quasi-hermitian}\footnote{The matrices $\M{A}$ and $\M{B}$ almost surely do not commute.}. Thus, at least one of the matrices should be positive definite (to serve as the non-trivial metric - see \eqref{intertwining-solution}). This observation should be kept in mind when reading \cite{BJN}. Diagrammatic derivation of the multiplication formula under the assumption that at least one of the matrices is positive (but not necessarily both) is a bit stronger than conventional free multiplication.  The latter assumes  {\em both} $\M{A}$ and $\M{B}$ are positive definite - a condition which guarantees commutativity of free multiplication. If neither of these matrices is positive definite, one has to double \eqref{BJN} and use the chiral $4N\times 4N$ hermitized form of $\M{Q}$, that is, 
\begin{equation}\label{hermitizedH}
\M{\tilde Q} = \left(\begin{array}{cc} \M{0} & \M{Q}\\ \M{Q}^\dagger & \M{0}\end{array}\!\!\right)   
\end{equation}
in order to compute the resolvent of $\M{AB}$ by means of planar diagrams \cite{FR}.

We shall now give a very basic physical interpretation of the trick of using \eqref{BJN} for diagrammatic derivation of the multiplication formula:
The hamiltonian governing small oscillations in our  system is 
\begin{equation}\label{hamiltonian}
{\cal H}  = \frac{1}{2} {\bf p}^T \frac{1}{\M{M}}{\bf p} + \frac{1}{2} {\bf x}^T \M{K}{\bf x}\,,
\end{equation}
leading to the equations of motion 
\begin{eqnarray}\label{hamiltonian-EOM}
\dot{\bf x} &=& {\partial {\cal H}\over \partial {\bf p}} = \M{M}^{-1}{\bf p}\nonumber\\
\dot{\bf p} &=& -{\partial {\cal H}\over \partial {\bf x}} = -\M{K}{\bf x}\,,
\end{eqnarray}
which are equivalent, of course, to \eqref{small-oscillations}. We can rewrite \eqref{hamiltonian-EOM} as  
\begin{equation}\label{hamiltonian-EOM1}
{d\over dt}\left(\begin{array}{c} {\bf x}\\{\bf p}\end{array}\right) = \M{\cal L}\left(\begin{array}{c} {\bf x}\\{\bf p}\end{array}\right)\,,
\end{equation}
where the constant matrix 
\begin{equation}\label{Liouvillian}
{\cal L} = \left(\begin{array}{cc} \M{0} & \M{M}^{-1}\\ \!\!\!\!-\M{K} & \M{0}\end{array}\!\!\right),   
\end{equation}
is the {\em Liouvillian} of our system. Thus, the solution of \eqref{hamiltonian-EOM1} is 
\begin{equation}\label{Liouvillian1}
\left(\begin{array}{c} {\bf x(t)}\\{\bf p}(t)\end{array}\right) = e^{\M{\cal L}t} \left(\begin{array}{c} {\bf x}(0)\\{\bf p}(0)\end{array}\right).
\end{equation}
Note that this hamiltonian time evolution preserves phase-space volume, ${\partial ({\bf x}(t), {\bf p}(t))\over \partial ({\bf x}(0), {\bf p}(0)) }= \det e^{\M{\cal L}t}  = e^{{\rm tr}\M{\cal L}} = 1$, in accordance with Liouville's theorem.

If the initial conditions in \eqref{Liouvillian1} coincide with one of the normal modes of the system with frequency $\omega$, clearly $$\left(\begin{array}{c} {\bf x(t)}\\{\bf p}(t)\end{array}\right) = e^{i\omega t} \left(\begin{array}{c} {\bf x}(0)\\{\bf p}(0)\end{array}\right).$$ Thus, the corresponding eigenvalue of $\M{{\cal L}}$ is $i\omega$. Another way to see this is to note that 
\begin{equation}\label{Liouvillian-square}
\M{\cal L}^2 = -\left(\begin{array}{cc} \M{H} &\!\! \M{0}\\ \!\!\! \M{0} &\M{H}^{\dagger }\end{array}\!\!\right)   
\end{equation}
and recall from \eqref{intertwining} that $\M{H}$ and $\M{H}^\dagger$ are similar to each other and have eigenvalues $\omega^2$. 

The Laplace transform of \eqref{Liouvillian1} involves the resolvent of $\M{{\cal L}}$. We readily obtain 
\begin{eqnarray}\label{Liouvillian-resolvent}
{1\over w-\M{\cal L}} &=&  \left(\begin{array}{cc} {w\over w^2 + \M{M}^{-1}\M{K}} & {\small\small\small\small\small{\M{M}^{-1}}}{1\over w^2 + \M{K} \M{M}^{-1}} \\ -{\small\small\small\small\small{\M{K}}} {1\over w^2 + \M{M}^{-1}\M{K}}& {w\over w^2 + \M{K} \M{M}^{-1}}\end{array}\!\!\right)\nonumber \\ &=&   \left(\begin{array}{cc} {w\over w^2 + \M{H}} & {\small\small\small\small\small{\M{M}^{-1}}}{1\over w^2 + \M{H}^\dagger} \\ -{\small\small\small\small\small{\M{K}}} {1\over w^2 + \M{H}}& {w\over w^2 + \M{H}^\dagger}\end{array}\!\!\right).
\end{eqnarray}
Let us now average \eqref{Liouvillian-resolvent} over $P_{\sigma_K}(\M{C}_1) $ and $P_{\sigma_M}(\M{C}_2) $ in \eqref{ginibre} and trace the four $N\times N$ blocks. The desired resolvent $G_{\M{H}}$ is obtained from the upper diagonal block 
\begin{equation}\label{diagonal-block}
w\Big\langle {1\over N}{\rm Tr} {1\over w^2 + \M{H}}\Big\rangle_{\sigma_M,\sigma_K} = -wG_{\M{H}}(-w^2;m_0,\sigma_M,\sigma_K).
\end{equation}
Thus, \eqref{BJN} can be thought of simply as the Liouvillian of some linearized hamiltonian system. 

\section{Statistical Mechanics of Phonons in the Random Matrix Model}\label{phonons} 
\begin{figure}[bt]
\includegraphics[width=\columnwidth]{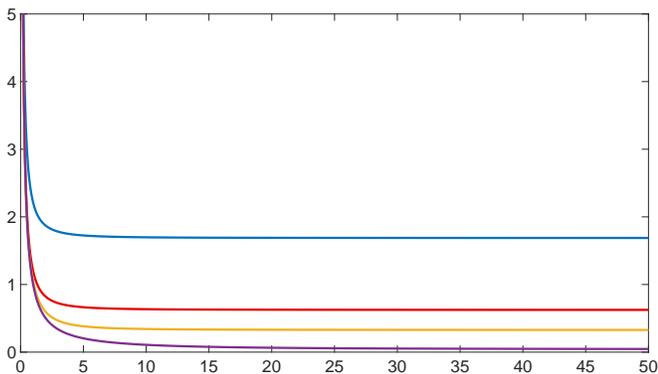}
\caption{Plot of the engergy density $u(T)$ given by \eqref{intensive} against inverse temperature $\beta$ (for $\hbar=1$). The parameters of the model are $\sigma_M=\sigma_K=1\, (\omega_0^2 = 1)$ and $\mu=10^{-4}$ (blue), $\mu=0.1$ (red) $\mu=1$ (yellow) and $\mu=100$ (purple).}\label{Fig:Energy}
\end{figure}

\begin{figure}[b]
\includegraphics[width=\columnwidth]{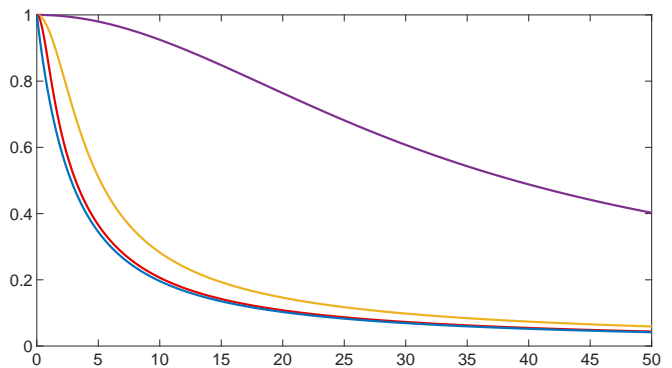}
\caption{Plot of the  the specific heat $c_V/k_\text{B}$ given by \eqref{intensive} against inverse temperature $\beta$ (for $\hbar=1$). The parameters and colors are as in Fig.~\ref{Fig:Energy}.}\label{Fig:CV}
\end{figure}

Upon quantization, the normal modes of our system amount to a collection of non-interacting quantum harmonic oscillators. In a state of thermal equilibrium at temperature $T$, the average energy tied with the oscillator with frequency $\omega$ is  
\begin{equation}\label{mode-energy}
\bar {\cal E}(\omega,T) = \hbar\omega\left(\frac{1}{2} + {1\over e^{\beta\hbar\omega}-1}\right)\,, 
\end{equation}
where $\beta = \frac{1}{k_\text{B} T}$, and the bar indicates thermal averaging with respect to the canonical density matrix 
\begin{equation}\label{canonical ensemble}
\hat\rho = 2\sinh\left(\frac{\beta\hbar\omega}{2}\right)\sum_{n=0}^\infty e^{-\beta\hbar\omega (n+\frac{1}{2})} |n\rangle\langle n|
\end{equation}
(written in the basis of oscillator energy eigenstates). The total average thermal energy $\bar E(T)$ of a given realization of our system of non-interacting oscillators, with eigenfrequencies $\omega_1, \omega_2,\ldots ,\omega_N$ is simply the sum of contributions of individual modes. Let 
\begin{equation}\label{mode-density}
\tilde\rho (\omega)  = \sum_{\alpha = 1}^N \delta(\omega-\omega_\alpha) = 2\omega\sum_{\alpha = 1}^N \delta(\omega^2-\omega_\alpha^2)
\end{equation}
be the density of modes (assuming $\omega>0$) in this realization. Thus, 
\begin{equation}\label{energy}
\bar E(T) = \int\limits_0^\infty \bar{\cal E}(\omega,T)\tilde\rho(\omega)d\omega\,.
\end{equation}
Finally, averaging over realizations in the random matrix ensemble, we obtain the ensemble average total thermal energy 
\begin{equation}\label{energy1}
\Big\langle \bar E(T)\Big\rangle_{\sigma_M,\sigma_K} = N\int\limits_0^\infty \bar{\cal E}(\omega,T)2\omega\varrho_{\M{H}}(\omega^2)d\omega\,,
\end{equation}
where we used the second equality in \eqref{mode-density} and \eqref{Hdensity}. This is, of course, an extensive quantity, proportional to $N$. Thus, the ensemble-averaged energy {\em density} and corresponding specific heat (per degree of freedom) are, respectively,  
\begin{eqnarray}\label{intensive}
u(T) &=& \frac{1}{N}\Big\langle \bar E(T)\Big\rangle_{\sigma_M,\sigma_K} = 2\int\limits_0^\infty \bar{\cal E}(\omega,T)\omega\varrho_{\M{H}}(\omega^2)d\omega\nonumber\\
c_V(T) &=& {\partial u(T)\over\partial T}\, =\frac{k_\text{B}(\beta \hbar)^2}{2} \! \int_0^\infty \!\! \frac{\omega^3 \varrho_{\M{H}}(\omega^2)}{\sinh^2(\beta\hbar\omega/2)} d\omega.\nonumber\\
\end{eqnarray}
We have carried numerical integration of \eqref{intensive} over the explicit expression for $\varrho_{\M{H}}(\omega^2)$ in Eq.~\eqref{eq:rho}, and compared them  to the results obtained by direct numerical averaging over realizations of the random matrix ensembles. The results are displayed in the Figs.~\ref{Fig:Energy} and \ref{Fig:CV} where we have only plotted the curves obtained from integration of the theoretical density for complex matrices (the relative errors compared to the other ones have been of order $10^{-6}$ for complex matrices and of $10^{-3}$ for real matrices). At large temperatures $c_V/k_\text{B}$ goes to 1 (the classical limit) while $u(T)$ goes to $k_\text{B} T$ (classical equipartition). High temperature means that $k_\text{B}T\gg \hbar \omega_\text{max}$ where $\omega_\text{max}=\sqrt{x_1}$. As $T\rightarrow 0$ all oscillation modes become frozen at their ground states with zero-point energy (ZPE) $\hbar \omega/2$. Thus the zero temperature limit of $u(T)$ is just the spectral sum over all ZPE up to exponentially small corrections $\sim e^{-\beta\hbar\omega}$. Consequentially $c_V$ is exponentially small and gets most of its contribution from the low frequency part of the spectrum. 
For large $\mu$ (see the purple plot in Fig.~\ref{Fig:CV}), as we discussed earlier, the spectral density tends to the Marchenko-Pastur profile, which has significant spectral weight at small frequencies. Thus, $c_V$ decays very slowly as a function of $\beta$ for large values of $\mu$.

{\it Summary.}
We have studied the vibration spectrum of highly connected systems. We have studied both concrete highly connected mechanical systems, namely, multi-segmented pendula, and also formulated a random matrix model describing such systems and analyzed it in detail. The latter analysis was achieved by employing $S$-transform techniques of free probability theory. Inside the bulk the analytical expression for the density of eigenvalues agrees well with simulations for finite $N$, but shows noticeable deviations at the edge, as it is typical for the large $N$ solution of the spectrum of random matrices. At the edge, numerics for the complex model shows an accurate fit with the Airy-kernel. Analytical treatment of this edge behavior is beyond the reaches of our approach and requires further investigation. Additional numerical results include the participation ratio of eigenvectors, which for the random matrix model shows that all vibration eigenmodes are extended, while there is a crossover from extended to localized eigenvectors for disordered pendula. Finally, based on our explicit analytical results for the density of eigenmodes, we have computed the thermodynamic properties of our matrix model in equilibrium. 

An important result of this paper is that the density of eigenfrequencies $\tilde\rho(\omega)$ of both our matrix model and pendula tend to a nonvanishing constant in the limit of small frequencies, which seems to be a common universal feature of highly connected systems. Low frequency modes are long-wavelength collective vibration modes, and they are expected to probe the mechanical system as a whole, in some sense. Thus, it is quite surprising that the density of these modes in our highly connected systems is qualitatively the same as that of acoustic phonons in one-dimensional perfect crystals, with its nearest-neighbor interatomic interactions, which is also flat constant. More formally, if we think of our ``hamiltonian" $\M{H}$ as the discrete laplacian of some graph associated with our highly connected mechanical system, then the {\it spectral dimension} $d_S$ of that graph is defined by the scaling behavior $\varrho_{\M{H}}(\omega^2)\sim (\omega^2)^{d_S/2-1}$ for $(\omega/\omega_0)^2 \ll 1$ \cite{HbA}, familiar from the theory of diffusion on fractal graphs \cite{AB,RT} (see also \cite{ADT}). Thus for our system, indeed $d_S=1$, which seems to be a {\it universal} feature of vibrational spectra of highly connected systems. Spectral dimension $d_S=2$ seems to correspond to the vibrational spectrum of globular proteins \cite{bA}. For a very recent discussion of spectral dimensions in the context of complex networks see \cite{Bianconi}.

This research was supported by the Israel Science Foundation (ISF) under grant No. 2040/17.  Computations presented in this work were performed on the Hive computer cluster at the University of Haifa, which is partly funded by ISF grant 2155/15. 

JF wishes to thank Andreas Fring, Tsampikos Kottos and Boris Shapiro for valuable discussions and for suggesting several references.

%\clearpage

\end{document}